\documentclass[preprint,prd,aps,showpacs,showkeys,nofootinbib]{revtex4}
\usepackage{graphicx}
\usepackage{amsmath} 
\usepackage{amssymb,xcolor}
\usepackage{color} 
\usepackage{subcaption}

\begin{document}
\title{Lepton flavor violating decays of Higgs boson in the NB-LSSM}
\author{Cai Guo$^{1,2,3}$,
	Xing-Xing Dong$^{1,2,3,4}$\footnote{dongxx@hbu.edu.cn},
	Shu-Min Zhao$^{1,2,3}$\footnote{zhaosm@hbu.edu.cn},\\
	Zhan Cao$^{1,2,3}$,
	Jia-Peng Huo$^{1,2,3,}$,
  Jin-Lei Yang$^{1,2,3}$,
  Tai-Fu Feng$^{1,2,3,5}$\footnote{fengtf@hbu.edu.cn}}	
\affiliation{
$^1$ College of Physics Science and Technology, Hebei University, Baoding, 071002, China\\
$^2$ Hebei Key Laboratory of High-precision Computation and Application of Quantum Field Theory, Baoding, 071002, China\\
$^3$ Hebei Research Center of the Basic Discipline for Computational Physics, Baoding, 071002, China\\
$^4$Departamento de F\'{i}sica and CFTP, Instituto Superior T\'{e}cnico, Universidade de Lisboa, Av. Rovisco Pais 1, 1049-001 Lisboa, Portugal\\
$^5$ Department of Physics, Chongqing University, Chongqing, 401331, China}

\begin{abstract}
 Lepton flavor violation (LFV) represents a clear new physics (NP) signal beyond the standard model (SM). NB-LSSM, the next to minimal supersymmetric extension of the SM with local B-L gauge symmetry, includes three Higgs singlets and three generations of right-handed neutrinos in the basis of MSSM, motivated by the new definition of SM-like Higgs resultly from the introducing of three Higgs singlets which mix with the two Higgs doublets at the tree level in the NB-LSSM. We calculate LFV processes $h\rightarrow l_i l_j$ in the mass eigenstate basis and the electroweak interaction basis separately, and the latter adopts the mass insertion approximation (MIA) method. In the suitable parameter space, we obtain the reasonable numerical results. At the same time, the corresponding constraints from the LFV rare decays $l_j\rightarrow l_i\gamma$ are considered to analyze the numerical results. 
  
\end{abstract}
\pacs{12.60.-i, 13.35.-r, 11.30.Hv} 
\keywords{Lepton flavor violation, Branching ratio, Beyond Standard Model}

\maketitle
\section{Introduction}
In the SM, the lepton-flavor number is conserved. However, the neutrino oscillation experiments\cite{neutrino1,neutrino2,neutrino3,neutrino4,neutrino5,neutrino7,neutrino8,neutrino9} have convinced that neutrinos possess tiny masses and mix with each other. Furthermore, the presentation of the GIM mechanism makes the LFV processes in the SM very tiny~\cite{CLFV1,CLFV2,CLFV3}, such as ${\rm{Br}}_{SM}(l_j\rightarrow l_i\gamma)\sim 10^{-55}$~\cite{SMljlig}. Therefore, the observations of LFV processes in future experiments indicate a definite evidence of NP beyond the SM. MEG Collaboration gives out the current experimental upper bound on the LFV process $\mu\rightarrow e \gamma$, which is ${\rm{Br}}(\mu\rightarrow e \gamma)< 4.2\times 10^{-13}$ at $90\%$ confidence level (C.L.)~\cite{MEG:2016leq}. The experimental results of Babar Collaboration indicate ${\rm{Br}}(\tau\rightarrow e \gamma)< 3.3\times 10^{-8}$ and ${\rm{Br}}(\tau\rightarrow \mu \gamma)< 4.2\times 10^{-8}$~\cite{BaBar:2009hkt}. Although these LFV processes are not observed so far, the parameter space of NP models suffers strict constraints from the corresponding experimental upper bounds.

The Higgs boson, an elementary particle, has been researched by the Large Hadron Collider (LHC) as one of the primary scientific goals. Combining the updated data of the ATLAS and CMS Collaborations, the measured mass of SM-like Higgs is $m_{h}=125.20\pm0.11 \rm {GeV}$~\cite{ATLAS:2023oaq}, which demonstrates that the Higgs mechanism is compelling. Recently, a search for LFV decays of the 125 GeV Higgs boson into the $e\mu$, $e\tau$, and $\mu\tau$ decay modes is presented. According to the latest experimental data provided by ATLAS and CMS, the observed upper limits on the LFV processes $h\rightarrow e\mu$, $h\rightarrow e\tau$, and $h\rightarrow \mu\tau$ at 95\% C.L. are~\cite{CMS:2023pte,ATLAS:2023mvd,CMS:2021rsq}
\begin{eqnarray}
	&&{\rm{Br}}(h\rightarrow e\mu)<4.4\times 10^{-5},\label{a1}\\
	&&{\rm{Br}}(h\rightarrow e\tau)<2.0\times 10^{-3},\label{a2}\\
	&&{\rm{Br}}(h\rightarrow \mu \tau)<1.5\times 10^{-3}.\label{a3}
\end{eqnarray} 

The LFV decays of Higgs boson are forbidden in the SM~\cite{Harnik}. But they can easily occur in NP models beyond the SM, for instance, the supersymmetric models~\cite{LFVHD3,ref-zhang-LFV,ref-zhang,YL19}, the composite Higgs model~\cite{LFVHD4,LFVHD5} and others~\cite{add9,add2}. In this work, we study the Higgs LFV decays $h\rightarrow l_i l_j$ in the next to minimum B-L supersymmetric model (NB-LSSM) ~\cite{Ahmed2021,Han2025,Barger2009}. Based on the minimum supersymmetric Standard Model (MSSM) ~\cite{Haber1985,Rosiek1990,Feng2009}, NB-LSSM extends the gauge symmetry group to \( SU(3)_C \otimes SU(2)_L \otimes U(1)_Y \otimes U(1)_{B-L} \), where B represents the baryon number and L stands for the lepton number. The invariance under \( U(1)_{B-L} \) gauge groups imposes R-parity conservation in the MSSM, which prevents proton decay~\cite{Aulakh1999}. The singlet scalar \( S \) can obtain a vacuum expectation value (VEV) \( \langle S \rangle = \frac{v_S}{\sqrt{2}} \sim \text{TeV} \) after breaking the local gauge symmetry, which is motivated to explain the \( \mu \) problem naturally. Besides, through the additional singlet Higgs states and right-handed (s)neutrinos, additional parameter space in the NB-LSSM is released from the LEP, Tevatron and LHC constraints to alleviate the hierarchy problem of the MSSM~\cite{Abdallah2017,Yang2020}. Besides, the NB-LSSM can also provide much more DM candidates~\cite{Khalil2009,Basso2012,Rose2017,Rose2018}. 

The most of researches about $h\rightarrow l_i l_j$ processes are studied with the mass eigenstate method. Using this method to find sensitive parameters is usually not intuitive and clear enough, which depends on the mass eigenstates of the particles and rotation matrices. It will lead us to pay too much attention to many unimportant parameters. To this end, we use the method called MIA to calculate these processes~\cite{Arganda2017,Wang2022,Moroi1996}. This method perturbatively treats the off-diagonal element in flavor entries $\Delta_{ij}^{XX}(X=L,R)$ of the slepton mass squared matrix in the electroweak basis, instead of diagonalizing the mass matrix in the physical basis. By means of mass insertions inside the propagators of the electroweak interaction eigenstates, at the analytical level, we can find many parameters that have direct impact on LFV. Compared with the mass eigenstate basis, the MIA provides a set of simple analytic formulas for the branch ratios of $h\rightarrow l_i l_j$. It can be emphasized which parameters will be effectively tested in the future colliders.

The outline of this paper is organized as follows. In section II, we present the ingredients of the NB-LSSM by introducing its superpotential, the general soft breaking terms, new corrected mass matrices and couplings. In section III, we explore the corresponding amplitudes and the branching ratios of rare LFV processes $l_j\rightarrow l_i \gamma$ and $h\rightarrow l_i l_j$ in the mass eigenstate basis and the electroweak interaction basis separately. Numerical results are tested against each other in two eigenstates and discussed in section IV. The conclusions are summarized in section V. The Feynman amplitude calculation of Fig.~\ref{fig1}(a), the one-loop functions, the needed Feynman rules in the electroweak interaction basis and the analytical expressions corresponding to Fig.~\ref{fig2} and Fig.~\ref{fig3} are collected in Appendix A, B, C and D, respectively.

\section{introduction of the NB-LSSM}
Using the local gauge group $U(1)_{B-L}$, we extend the MSSM to obtain the NB-LSSM with the local gauge group $SU(3)_C\times SU(2)_L \times U(1)_Y\times U(1)_{B-L}$. Because of the introduction of three Higgs singlets, the Higgs mass squared matrix is 5×5. This can not only explain the 125GeV Higgs mass easily, but also enrich the Higgs physics.

In Table~\ref{quarks}, we show the chiral superfields and quantum numbers of the NB-LSSM. The corresponding superpotential of the NB-LSSM is shown as
\begin{eqnarray}
	&&W=-Y_d\hat{d}\hat{q}\hat{H}_d-Y_e\hat{e}\hat{l}\hat{H}_d-\lambda_2\hat{S}\hat{\chi}_1\hat{\chi}_2+\lambda\hat{S}\hat{H}_u\hat{H}_d+\frac{\kappa}{3}\hat{S}\hat{S}\hat{S}+Y_u\hat{u}\hat{q}\hat{H}_u+Y_{\chi}\hat{\nu}\hat{\chi}_1\hat{\nu}
	\nonumber\\&&~~~~~~~+Y_\nu\hat{\nu}\hat{l}\hat{H}_u.
\end{eqnarray}
Here, $\hat{\chi}_1,~\hat{\chi}_2,~\hat{S}$ are three Higgs singlets. $Y_{u,d,e,\nu,\chi}$ are the Yukawa couplings. $\lambda$, $\lambda_2$ and $\kappa$ are the dimensionless couplings. These couplings can produce one-loop diagrams influencing the LFV decays $h\rightarrow l_i l_j$, which enrich the lepton physics in a certain degree. It is important to note that the term \(Y_{\nu}^{\prime} \hat{\nu} \hat{l} \hat{S}\) is not allowed, as the sum of \(U(1)_{Y}\) charges of \(\hat{\nu}, \hat{l}, \hat{S}\) does not satisfy the necessary charge neutrality condition.

The soft SUSY breaking terms are
\begin{eqnarray}
	&&\mathcal{L}_{soft}=\mathcal{L}_{soft}^{MSSM}-\frac{T_\kappa}{3}S^3+T_{\lambda}SH_d^iH_u^j+T_{2}S\chi_1\chi_2\nonumber\\&&
	-T_{\chi,ik}\chi_1\tilde{\nu}_{R,i}^{*}\tilde{\nu}_{R,k}^{*}
	-T_{\nu,ij}H_u^i\tilde{\nu}_{R,i}^{*}\tilde{e}_{L,j}-m_{\eta}^2|\chi_1|^2-m_{\bar{\eta}}^2|\chi_2|^2\nonumber\\&&-m_S^2|S|^2-m_{\nu,ij}^2\tilde{\nu}_{R,i}^{*}\tilde{\nu}_{R,j}
	-\frac{1}{2}(2M_{BB^\prime}\tilde{B}\tilde{B^\prime}+M_{BL}\tilde{B^\prime}^2+h.c.).
\end{eqnarray}
$\mathcal{L}_{soft}^{MSSM}$ represents the soft breaking terms in the MSSM. $T_{\kappa}$, $T_{\lambda}$, $T_2$, $T_{\chi}$ and $T_{\nu}$ are all trilinear coupling coefficients. For the soft breaking slepton mass matrices $m^{2}_{\tilde{L}, \tilde{E}}$ and the trilinear coupling matrix $T_{e}$, we introduce the slepton flavor mixings, which take into account the off-diagonal terms~\cite{Arganda2016,Zhang2014,Calibbi2018}. These mixings are parametrized by means of a complete set of slepton flavor mixing dimensionless parameters $\delta^{XX}_{ij}$ with $XX = LL, RR, LR(RL=LR)$~\cite{Arganda2016}, and flavor indices $i, j = 1, 2, 3$, with $i \neq j$,
\begin{eqnarray}
	m_{\tilde{L}}^2\hspace{-0.1cm}=\hspace{-0.2cm}\left(\hspace{-0.1cm}\begin{array}{ccc}
		m_{L}^2 & \delta_{12}^{LL}m_{LL}^2 &  \delta_{13}^{LL}m_{LL}^2\\
		\delta_{12}^{LL}m_{LL}^2 & m_{L}^2 & \delta_{23}^{LL}m_{LL}^2\\
		\delta_{13}^{LL}m_{LL}^2 & \delta_{23}^{LL}m_{LL}^2 &m_{L}^2
	\end{array}\hspace{-0.1cm}\right)\hspace{-0.1cm},~~~
\end{eqnarray}
\begin{align}
	m_{\tilde{E}}^2 &= \begin{pmatrix}
		m_{E}^2 & \delta_{12}^{RR}m_{EE}^2 &  \delta_{13}^{RR}m_{EE}^2 \\
		\delta_{12}^{RR}m_{EE}^2 & m_{E}^2 & \delta_{23}^{RR}m_{EE}^2 \\
		\delta_{13}^{RR}m_{EE}^2 & \delta_{23}^{RR}m_{EE}^2 & m_{E}^2
	\end{pmatrix}, \label{eq:mass_matrix} \\
	T_e &= \begin{pmatrix}
		1 & \delta_{12}^{LR} &  \delta_{13}^{LR} \\
		\delta_{12}^{LR} & 1 & \delta_{23}^{LR} \\
		\delta_{13}^{LR} & \delta_{23}^{LR} & 1
	\end{pmatrix} A_e. \label{eq:transformation_matrix}
\end{align}
 We show the concrete forms of the two Higgs doublets and three Higgs singlets
\begin{eqnarray}
	&&\hspace{-2cm}H^0_d=\frac{1 }{\sqrt{2}} \phi_{d}+{\frac{1 }{\sqrt{2}} }v_{d}+i{\frac{1 }{\sqrt{2}} }\sigma_d,
	\nonumber\\&&\hspace{-2cm}H^0_u={\frac{1 }{\sqrt{2}} }\phi_{u}+{\frac{1 }{\sqrt{2}} }v_{u}+i{\frac{1 }{\sqrt{2}} }\sigma_u,
	\nonumber\\&&\hspace{-2cm}\chi_1={\frac{1 }{\sqrt{2}} }\phi_{1}+{\frac{1 }{\sqrt{2}} }v_{\eta}+i{\frac{1 }{\sqrt{2}} }\sigma_1,
	\nonumber\\&&\hspace{-2cm}\chi_2={\frac{1 }{\sqrt{2}} }\phi_{2}+{\frac{1 }{\sqrt{2}} }v_{\bar{\eta}}+i{\frac{1 }{\sqrt{2}} }\sigma_2,
	\nonumber\\&&\hspace{-2cm}S={\frac{1 }{\sqrt{2}} }\phi_S+{\frac{1 }{\sqrt{2}} }v_S+i{\frac{1 }{\sqrt{2}} }\sigma_S.
\end{eqnarray}
The vacuum expectation VEVs of the Higgs superfields $H_u$, $H_d$, $\chi_1$, $\chi_2$ and $S$ are presented by
$v_u,~v_d,~v_\eta,~v_{\bar\eta}$ and $v_S$ respectively. Two angles are defined as
$\tan\beta=v_u/v_d$ and $\tan\beta^{\prime }=v_{\bar{\eta}}/v_{\eta}$.
\begin{table}[!h]
	\caption{ The chiral superfields and quantum numbers in NB-LSSM}
	\begin{tabular}{|c|c|c|c|c|}
		\hline
		Superfields & $U(1)_Y$ & $SU(2)_L$ & $SU(3)_C$ & $U(1)_{B-L}$ \\
		\hline
		$\hat{q}$ & 1/6 & 2 & 3 & 1/6  \\
		\hline
		$\hat{l}$ & -1/2 & 2 & 1 & -1/2  \\
		\hline
		$\hat{H}_d$ & -1/2 & 2 & 1 & 0 \\
		\hline
		$\hat{H}_u$ & 1/2 & 2 & 1 & 0 \\
		\hline
		$\hat{d}$ & 1/3 & 1 & $\bar{3}$ & -1/6  \\
		\hline
		$\hat{u}$ & -2/3 & 1 & $\bar{3}$ & -1/6 \\
		\hline
		$\hat{e}$ & 1 & 1 & 1 & $1/2$  \\
		\hline
		$\hat{\nu}$ & 0 & 1 & 1 & $1/2$ \\
		\hline
		$\hat{\chi}_1$ & 0 & 1 & 1 & -1 \\
		\hline
		$\hat{\chi}_2$ & 0 & 1 & 1 & 1\\
		\hline
		$\hat{S}$ & 0 & 1 & 1 & 0 \\
		\hline
	\end{tabular}
	\label{quarks}
\end{table}

$U(1)_Y$ and $U(1)_{B-L}$ have the gauge kinetic mixing, which can also be induced through RGEs even with zero value at $M_{GUT}$. The covariant derivatives of this model can be written as
\begin{eqnarray}
	&&D_\mu=\partial_\mu-i\left(\begin{array}{cc}Y,&B-L\end{array}\right)
	\left(\begin{array}{cc}g_{Y} &g{'}_{{YB}}\\g{'}_{{BY}} &g{'}_{{B-L}}\end{array}\right)
	\left(\begin{array}{c}A_{\mu}^{\prime Y} \\ A_{\mu}^{\prime BL}\end{array}\right)\;,
	\label{gauge1}
\end{eqnarray}
where $Y$ and $B-L$ represent the hypercharge and $B-L$ charge, respectively. The two Abelian gauge groups are unbroken, then the change of basis can occur with the rotation matrix $R$ ($R^TR=1$)~\cite{Belanger2017,Barger2009,Chankowski2006,Yang2018},
\begin{eqnarray}
	&&\left(\begin{array}{cc}g_{Y} &g{'}_{{YB}}\\g{'}_{{BY}} &g{'}_{{B-L}}\end{array}\right)
	R^T=\left(\begin{array}{cc}g_{1} &g_{{YB}}\\0 &g_{{B}}\end{array}\right)\;.
	\label{gauge3}
\end{eqnarray}
As a result, the $U(1)$ gauge fields are redefined as
\begin{eqnarray}
	&&R\left(\begin{array}{c}A_{\mu}^{\prime Y} \\ A_{\mu}^{\prime BL}\end{array}\right)
	=\left(\begin{array}{c}A_{\mu}^{Y} \\ A_{\mu}^{BL}\end{array}\right)\;.
	\label{gauge4}
\end{eqnarray}

In the NB-LSSM, four two-component spinors ($\tilde{W}^{-}$, $\tilde{W}^{+}$, $\tilde{H}^{-}$, $\tilde{H}^{+}$) form two four-component Dirac fermions (Charginos) $\chi^{+}$, $\chi^{-}$
\begin{eqnarray}
	M_{\chi^{\pm}}=\begin{pmatrix}
		M_{2} & \frac{1}{\sqrt{2}}g_{2}v_{u} \\
		\frac{1}{\sqrt{2}}g_{2}v_{d} & \frac{1}{\sqrt{2}} \lambda v_S 
	\end{pmatrix}.
\end{eqnarray}
This matrix is diagonalized by $U$ and $V$:
\begin{equation}
	U^{*} M_{\chi^{\pm}} V^{\dagger}=M_{\chi^{\pm}}^{\text {diag }}.
\end{equation}

The mass matrix for neutralino in the basis $(\tilde{B}, \tilde{W}^3, \tilde{H}_1, \tilde{H}_2,
\tilde{B'}, \tilde{\chi_1}, \tilde{\chi_2}, \tilde{S})$ is

\begin{equation}
	m_{\chi^0} = \left(
	\begin{array}{cccccccc}
		M_1 &0 &-\frac{1}{2}g_1 v_d &\frac{1}{2} g_1 v_u &{M}_{B B'} & 0  & 0  &0\\
		0 &M_2 &\frac{1}{2} g_2 v_d  &-\frac{1}{2} g_2 v_u  &0 &0 &0 &0\\
		-\frac{1}{2}g_1 v_d &\frac{1}{2} g_2 v_d  &0
		&- \frac{1}{\sqrt{2}} {\lambda} v_S&-\frac{1}{2} g_{YB} v_d &0 &0 & - \frac{1}{\sqrt{2}} {\lambda} v_u\\
		\frac{1}{2}g_1 v_u &-\frac{1}{2} g_2 v_u  &- \frac{1}{\sqrt{2}} {\lambda} v_S &0 &\frac{1}{2} g_{YB} v_u  &0 &0 &- \frac{1}{\sqrt{2}} {\lambda} v_d\\
		{M}_{B B'} &0 &-\frac{1}{2} g_{YB} v_{d}  &\frac{1}{2} g_{YB} v_{u} &{M}_{BL} &- g_{B} v_{\eta}  &g_{B} v_{\bar{\eta}}  &0\\
		0  &0 &0 &0 &- g_{B} v_{\eta}  &0 &-\frac{1}{\sqrt{2}} {\lambda}_{2} v_S  &-\frac{1}{\sqrt{2}} {\lambda}_{2} v_{\bar{\eta}} \\
		0 &0 &0 &0 &g_{B} v_{\bar{\eta}}  &-\frac{1}{\sqrt{2}} {\lambda}_{2} v_S  &0 &-\frac{1}{\sqrt{2}} {\lambda}_{2} v_{\eta} \\
		0 &0 & - \frac{1}{\sqrt{2}} {\lambda} v_u &- \frac{1}{\sqrt{2}}{\lambda} v_d &0 &-\frac{1}{\sqrt{2}} {\lambda}_{2} v_{\bar{\eta}}
		&-\frac{1}{\sqrt{2}} {\lambda}_{2} v_{\eta}  &\sqrt{2}\kappa v_S\end{array}
	\right).\label{neutralino}
\end{equation}
This matrix is diagonalized by the rotation matrix $N$,
\begin{equation}
	N^{*}m_{\chi^0} N^{\dagger} = m^{diag}_{\chi^0}.
\end{equation}

Based on $(\tilde{e}_{L}, \tilde{e}_{R})$, the mass squared matrix for sleptons reads
\begin{eqnarray}
	&&\tilde{M}^{2}_{\tilde{l}}=\begin{pmatrix}
		(M^{2}_{L})_{LL} & (M^{2}_{L})_{LR} \\
		(M^{2}_{L})^{\dagger}_{LR} & (M^{2}_{L})_{RR}
	\end{pmatrix},
\end{eqnarray}
\begin{eqnarray}
	&&(M^{2}_{L})_{LL}=m^{2}_{\tilde{L}}+\mathbf{1}\frac{1}{8}\big((g^{2}_{1}+g^{2}_{YB})(-v^{2}_{u}+v^{2}_{d})+g^{2}_{2}(-v^{2}_{d}+v^{2}_{u})+g_{YB}g_{B}(-2v^{2}_{\bar{\eta}}+2v^{2}_{\eta}-v^{2}_{d} \nonumber\\
	&&\;\;\;\;\;\;\;\;\;\;\;\;\;\;\;\;\; +v^{2}_{u})+2g^{2}_{B}(-v^{2}_{\bar{\eta}}+v^{2}_{\eta})\big)+\frac{1}{2}v^{2}_{d}Y^{\dagger}_{e}Y_{e}, \nonumber\\
	&&(M^{2}_{L})_{RR}=m^{2}_{\tilde{E}}+\mathbf{1}\frac{1}{8}\big(2(g^{2}_{1}+g^{2}_{YB})(-v^{2}_{d}+v^{2}_{u})-2g^{2}_{B}(-v^{2}_{\bar{\eta}}+v^{2}_{\eta})+g_{B}g_{YB}(4v^{2}_{\bar{\eta}}-4v^{2}_{\eta}-v^{2}_{d} \nonumber\\
	&&\;\;\;\;\;\;\;\;\;\;\;\;\;\;\;\;\;  +v^{2}_{u})\big)+\frac{1}{2}v^{2}_{d}Y_{e}Y^{\dagger}_{e}, \nonumber\\
	&&(M^{2}_{L})_{LR}=\frac{1}{\sqrt{2}}(v_{d}T_{e}^{\dagger}-v_{u}\frac{\lambda \upsilon_S }{2}  Y^{\dagger}_{e}).
	\label{eq111}
\end{eqnarray}
This matrix is diagonalized by $Z^{E}$:\begin{eqnarray}
	&&Z^{E} \tilde{M}^{2}_{\tilde{l}} Z^{E,\dagger}=({M^{2}_{\tilde{l}}})^{\text {diag }}.
\end{eqnarray}

The mass matrix for Higgs in the basis $(\phi_d,\phi_u,\phi_1,\phi_2,\phi_S) $ is

\begin{equation}
	m_{\tilde{h}} ^2 = \left(
	\begin{array}{cccccccc}
		m_{\phi_d} m_{\phi_d}  &m_{\phi_u} m_{\phi_d} &m_{\phi_1} m_{\phi_d} &m_{\phi_2} m_{\phi_d} &m_{\phi_S} m_{\phi_d}\\
		m_{\phi_d} m_{\phi_u}  &m_{\phi_u} m_{\phi_u} &m_{\phi_1} m_{\phi_u} &m_{\phi_2} m_{\phi_u} &m_{\phi_S} m_{\phi_u}\\
		m_{\phi_d} m_{\phi_1}  &m_{\phi_u} m_{\phi_1} &m_{\phi_1} m_{\phi_1} &m_{\phi_2} m_{\phi_1} &m_{\phi_S} m_{\phi_1}\\
		m_{\phi_d} m_{\phi_2}  &m_{\phi_u} m_{\phi_2} &m_{\phi_1} m_{\phi_2} &m_{\phi_2} m_{\phi_2} &m_{\phi_S} m_{\phi_2}\\
		m_{\phi_d} m_{\phi_S}  &m_{\phi_u} m_{\phi_S} &m_{\phi_1} m_{\phi_S} &m_{\phi_2} m_{\phi_S} &m_{\phi_S} m_{\phi_S}\\ 
	\end{array}
	\right).\label{higgs}
\end{equation}

\begin{align}    
	m_{\phi_d} m_{\phi_d} &= \frac{1}{4} G^2 v_d^2 - \frac{1}{4} \left( -2\sqrt{2} v_S \Re(T_\lambda)+ (-\kappa v_S^2 + \lambda_2 v_{\eta} v_{\bar{\eta}}  )\lambda^* + \lambda(v_{\eta} v_{\bar{\eta}} \lambda_2^* - v_S^2 \kappa^*) \right) \tan\beta \nonumber\\   
	m_{\phi_d} m_{\phi_u} &=-\frac{1}{4} G^2 v_d v_u + \frac{1}{4} \left( -2 \sqrt{2} v_S \Re(T_\lambda) + (4\lambda v_d v_u - \kappa v_S^2 + \lambda_2 v_{\eta} v_{\bar{\eta}})\lambda^* + \lambda (v_{\eta} v_{\bar{\eta}} \lambda_2^* - v_S^2 \kappa^*) \right) \nonumber\\    
	m_{\phi_u} m_{\phi_u} &=\frac{1}{4} G^2 v_u^2 - \frac{1}{4} \left( -2\sqrt{2} v_S \Re(T_\lambda) + (-\kappa v_S^2 + \lambda_2 v_{\eta} v_{\bar{\eta}})\lambda^* + \lambda(v_{\eta} v_{\bar{\eta}} \lambda_2^* - v_S^2 \kappa^*) \right) \cot\beta \nonumber\\  
	m_{\phi_d} m_{\phi_1} &= \frac{1}{2}(g_1 + g_{YB} g_B) v_d v_{\eta} + \frac{1}{2} v_u  v_{\bar{\eta}} \Re(\lambda^* \lambda_2) \nonumber\\    
	m_{\phi_u} m_{\phi_1} &= -\frac{1}{2}(g_1 +g_{YB} g_B) v_u v_{\eta} + \frac{1}{2} v_d v_{\bar{\eta}} \Re(\lambda^* \lambda_2) \nonumber\\  
	m_{\phi_1} m_{\phi_1} &= g_B^2 v_{\eta}^2 - \frac{1}{4} \left( -2\sqrt{2} v_S \Re(T_2)+ (-\kappa v_S^2 + \lambda v_d v_u)\lambda_2^*+ \lambda_2 (v_d v_u \lambda^* - v_S^2 \kappa^*) \right) \tan\beta' \nonumber\\ 
	m_{\phi_d} m_{\phi_2} &= -\frac{1}{2}(g_{YB} g_B) v_d v_{\bar{\eta}} + \frac{1}{2} v_u v_{\eta} \Re(\lambda^* \lambda_2) \nonumber\\    
	m_{\phi_u} m_{\phi_2} &= \frac{1}{2}(g_{YB} g_B) v_u v_{\bar{\eta}}+ \frac{1}{2} v_d v_{\eta} \Re(\lambda^* \lambda_2) \nonumber\\    
	m_{\phi_1} m_{\phi_2} &= \frac{1}{4}\left(-2 \sqrt{2} v_S \Re(T_2) +\left(4\lambda_2 v_{\eta} v_{\bar{\eta}}-\kappa v_S^2 + \lambda v_d v_u\right)\lambda_2^* + \lambda_2\left(v_d v_u \lambda^* - v_S^2\kappa^*\right) \right)-g_B^2 v_{\eta} v_{\bar{\eta}} \nonumber\\
	m_{\phi_2} m_{\phi_2} &= g_B^2 v_{\bar{\eta}}^2 - \frac{1}{4} \left( -2\sqrt{2} v_S \Re(T_2) + (-\kappa v_S^2 + \lambda v_d v_u)\lambda_2^* + \lambda_2(v_d v_u \lambda^* - v_S^2 \kappa^*) \right) \cot\beta' \nonumber\\  
	m_{\phi_d} m_{\phi_S} &= \frac{1}{2} \left( v_S \left( 2 \lambda v_d - \kappa v_u \right) \lambda - v_u \left( \lambda v_S \kappa + \sqrt{2} \Re(T_\lambda) \right) \right) \nonumber\\
	m_{\phi_u} m_{\phi_S} &= \frac{1}{2} \left( -v_d \left( \lambda v_S \kappa^*+ \sqrt{2}  \Re(T_\lambda) \right) - v_S \left( -2 \lambda v_u + \kappa v_d \right) \lambda^* \right)\nonumber\\
	m_{\phi_1} m_{\phi_S} &= \frac{1}{2} \left( -v_{\bar{\eta}} \left( \lambda_2 v_S \kappa^*+ \sqrt{2}  \Re(T_2) \right) + v_S \left( 2 \lambda_2 v_{\eta}-\kappa v_{\bar{\eta}} \right) \lambda_2^* \right)\nonumber\\
	m_{\phi_2} m_{\phi_S} &= \frac{1}{2} \left( -v_{\eta} \left( \lambda_2 v_S \kappa^*+ \sqrt{2}  \Re(T_2) \right) - v_S \left( -2 \lambda_2 v_{\bar{\eta}}+\kappa v_{\eta} \right) \lambda_2^* \right)\nonumber\\
    m_{\phi_S} m_{\phi_S} &= 2 \kappa^* \kappa v_S^2 + \frac{\sqrt{2}}{2} \Re(T_\kappa) v_S + \frac{\sqrt{2}}{2} \Re(T_\lambda) \frac{v_u v_d}{v_S} + \frac{\sqrt{2}}{2} \Re(T_2) \frac{v_{\eta}v_{\bar{\eta}}} {v_S}\
    \label{21}
\end{align}
This matrix is diagonalized by the rotation matrix $Z^H$,
\begin{equation}
	Z^H m_{\tilde{h}}^2 Z^{H \dagger} = (m_{h_i^0}^2)^{diag}.
\end{equation}
Here, $G^2=g_{1}^2+g_{2}^2+g_{{YB}}^2$. The mass of the SM-like Higgs boson can be obtained after considering the leading-log radiative corrections from stop and top particles.
\begin{eqnarray}
	m_{h}=\sqrt{(m_{h_1^0})^2+\Delta m_h^2},
\end{eqnarray}
where $m_{h_1^0}$ represents the lightest tree-level Higgs boson mass, and the leading-log radiative corrections $\Delta m_h^2$ can be written as~\cite{HiggsC1,HiggsC3}
\begin{eqnarray}
	&&\Delta m_h^2=\frac{3m_t^4}{4 \pi ^2 v^2}\Big[\Big(\tilde{t}+\frac{1}{2}\tilde{X}_t\Big)+\frac{1}{16 \pi ^2}\Big(\frac{3m_t^2}{2v^2}-32\pi\alpha_3\Big)(\tilde{t}^2+\tilde{X}_t\tilde{t})\Big],\nonumber \\&&\tilde{t}=\log\frac{M_S^2}{m_t^2},~~~~\tilde{X}_t=\frac{2\tilde{A}_t^2}{M_S^2}(1-\frac{\tilde{A}_t^2}{12M_S^2}).
\end{eqnarray}
Here, $\alpha_3$ is the strong coupling constant, $M_S=\sqrt{m_{\tilde{t}_1}m_{\tilde{t}_2}}$ with $m_{\tilde{t}_{1,2}}$ are the stop masses, $\tilde{A}_t=A_t-\frac{\lambda v_S }{\sqrt{2}}\cot \beta $ with $A_t=T_{u,33}$ denotes the trilinear Higgs-stops coupling.

\section{The processes $l_j\rightarrow l_i \gamma$ and $h\rightarrow l_i l_j$ in the NB-LSSM}
In this section, we analyze the branching ratios of LFV processes $l_j\rightarrow l_i \gamma$ and $h\rightarrow l_i l_j$ in the mass eigenstate basis and the electroweak interaction basis, the latter adopt the MIA method. The concrete contents will be discussed as follows.

\subsection{Rare decays $l_j\rightarrow l_i \gamma$}
\begin{figure}[t]
\centering
\includegraphics[width=12cm]{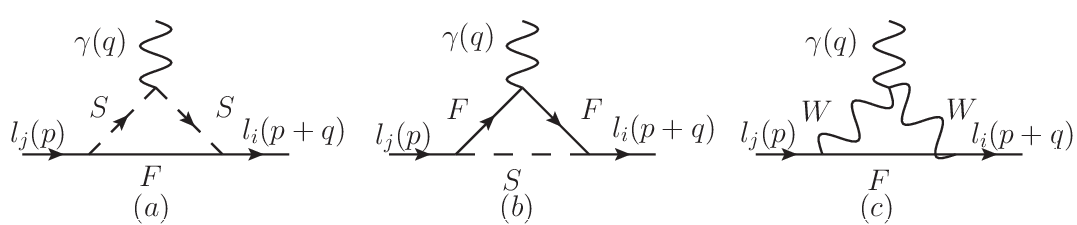}\\
\caption{The triangle type diagrams for decays $l_j\rightarrow l_i \gamma$.} \label{fig1}
\end{figure}

Generally, the corresponding effective amplitude for processes $l_j\rightarrow l_i\gamma$ can be written as~\cite{efflj}
\begin{eqnarray}
&&{\cal M}=e\epsilon^{\mu}{\bar u}_i(p+q)[q^2\gamma_{\mu}(D_1^LP_L+D_1^RP_R)+m_{l_j}i\sigma_{\mu\nu}q^{\nu}(D_2^LP_L+D_2^RP_R)]u_j(p),
\nonumber\\&&D_{\alpha}^{L,R}=D_{\alpha}^{L,R}(n)+D_{\alpha}^{L,R}(c)+D_{\alpha}^{L,R}(W),\alpha=1,2,
\end{eqnarray}
where $p$ ($q$) represents the injecting lepton (photon) momentum. $m_{l_j}$ is the j-th generation lepton mass. $\epsilon$ is the photon polarization vector and $u_i(p)$ ($v_i(p)$) is the lepton (antilepton) wave function. In Fig.~\ref{fig1}, we show the relevant Feynman diagrams corresponding to above amplitude. The Wilson coefficients $D_{\alpha}^{L,R}(\alpha=1,2)$ are determined by summing the amplitudes of the corresponding diagrams and discussed as follows.

$D_{\alpha}^{L,R}(n)(\alpha=1,2)$, the virtual neutral fermion contributions corresponding to Fig.~\ref{fig1}(a) are derived as follows. For clarity, the main text presents only key results, while the complete derivation is provided in Appendix~\ref{calculation process}.
\begin{eqnarray}
&&D_1^L(n)=\sum_{F=\chi_k^0}\sum_{S=\tilde{l}}\frac{1}{6{\Lambda}^2}
H_R^{SF\bar{l}_i}H_L^{S^*l_j\bar{F}}I_1(x_F,x_S),
\nonumber\\&&D_2^L(n)=\sum_{F=\chi^0_k}\sum_{S=\tilde{l}}\frac{m_F}{m_{l_j}{\Lambda}^2}
H_L^{SF\bar{l}_i}H_L^{S^*l_j\bar{F}}[I_2(x_F,x_S)-I_3(x_F,x_S)],
\nonumber\\&&D_{\alpha}^R(n)=D_{\alpha}^L(n)|_{L{\leftrightarrow}R},\alpha=1,2,\label{26}
\end{eqnarray}
where $x_i=m_i^2/{\Lambda}^2$, $m_i$ is the corresponding particle mass and ${\Lambda}$ is the NP energy scale. $H_{L,R}^{SF\bar{l}_i}$ represent the left (right)-hand part of the coupling vertex.
Then, we discuss the virtual charged fermion contributions $D_{\alpha}^{L,R}(c)(\alpha=1,2)$ corresponding to Fig.~\ref{fig1}(b)
\begin{eqnarray}
	&&D_1^L(c)=\sum_{F=\chi^{\pm}}\sum_{S=\tilde{\nu}^{I(R)}}\frac{1}{6{\Lambda}^2}
	H_R^{SF\bar{l}_i}H_L^{S^*l_j\bar{F}}[3I_2(x_S,x_F)-I_4(x_S,x_F)],
	\nonumber\\&&D_2^L(c)=\sum_{F=\chi^{\pm}}\sum_{S=\tilde{\nu}^{I(R)}}\frac{m_F}{m_{l_j}{\Lambda}^2}
	H_L^{SF\bar{l}_i}H_L^{S^*l_j\bar{F}}I_2(x_S,x_F),
	\nonumber\\&&D_{\alpha}^R(c)=D_{\alpha}^L(c)|_{L{\leftrightarrow}R},\alpha=1,2
\end{eqnarray}
 The concrete expressions for one-loop functions $I_i(x_1,x_2)(i=1,2,3,4)$ are collected in Appendix~\ref{OLF}. The contributions from $W$-$W$-neutrino diagram can be ignored due to the tiny neutrino mass. 

We deduce the decay widths for processes $l_j\rightarrow l_i \gamma$ as
\begin{eqnarray}
&&\Gamma\left(l_j\rightarrow l_i\gamma\right)=\frac{e^2}{16\pi}m_{l_j}^{5}\left(|D_2^L|^2+|D_2^R|^2\right).
\end{eqnarray}
Then, the concrete branching ratios of $l_j\rightarrow l_i \gamma$ can be expressed as~\cite{efflj}
\begin{eqnarray}
&&Br\left(l_j\rightarrow l_i\gamma\right)=\Gamma\left(l_j\rightarrow l_i\gamma\right)/\Gamma_{l_j}.
\end{eqnarray}
Here, $\Gamma_{l_j}$ represent the total decay widths of the charged leptons $l_j$. We take $\Gamma_{\mu}\simeq2.996\times10^{-19}$ GeV and $\Gamma_{\tau}\simeq2.265\times10^{-12}$ GeV~\cite{HCSWG} in our latter numerical calculations.
\subsection{Rare decay $h\rightarrow l_i l_j$}

\begin{figure}[t]
	\centering
	\includegraphics[width=14cm]{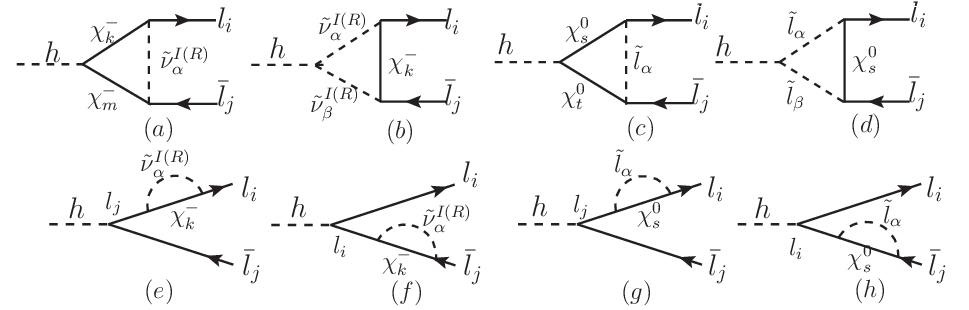}\\
	\caption{The triangle and self-energy type diagrams for decays $h\rightarrow l_i\bar{l}_j $.} \label{fig2}
\end{figure}

The corresponding effective amplitude for $h\rightarrow l_i\bar{l}_j $ can be summarized as
\begin{eqnarray}
&&{\cal M}={\bar u}_i(q)(K_LP_L+K_RP_R)v_j(p),
\nonumber\\&&K_{L,R}=K_{L,R}(S_1)+K_{L,R}(S_2)+K_{L,R}(S_3).
\end{eqnarray}
Here $K_{L,R}(S_1)$ are the coupling coefficients corresponding to triangle diagrams in Fig.~\ref{fig2}(a) and Fig.~\ref{fig2}(c), $K_{L,R}(S_2)$ denote the contributions to triangle diagrams in Fig.~\ref{fig2}(b) and Fig.~\ref{fig2}(d). $K_{L,R}(S_3)$ denote the contributions to self-energy diagrams from Fig.~\ref{fig2}(e) to Fig.~\ref{fig2}(h). We give out the concrete expressions for these contributions as follows.

The contributions from triangle diagrams in Fig.~\ref{fig2}:
\begin{eqnarray}
&&K_L(S_1)=\sum_{F=\chi^{\pm},\chi^0}\sum_{S=\tilde{\nu}^{I(R)},\tilde{l}}
\frac{m_F}{{\Lambda}^2}H_L^{S_2F\bar{l}_i}H^{hS_1S_2^*}H_L^{S_1^*l_j\bar{F}}G_1(x_F,x_{S_1},x_{S_2}),
\nonumber\\&&K_R(S_1)=K_L(S_1)|_{L{\leftrightarrow}R},\nonumber\\
&&K_L(S_2)=\sum_{F=\chi^{\pm},\chi^0}\sum_{S=\tilde{\nu}^{I(R)},\tilde{l}}
\big[H_L^{SF_2\bar{l}_i}H_R^{hF_1\bar{F}_2}H_L^{S^*l_j\bar{F}_1}G_2(x_S,x_{F_1},x_{F_2})
\nonumber\\&&\hspace{1.5cm}+\frac{m_{F_1}m_{F_2}}{{\Lambda}^2}
H_L^{SF_2\bar{l}_i}H_L^{hF_1\bar{F}_2}H_L^{S^*l_j\bar{F}_1}G_1(x_S,x_{F_1},x_{F_2})\big],
\nonumber\\&&K_R(S_2)=K_L(S_2)|_{L{\leftrightarrow}R}.
\end{eqnarray}

The contributions from self-energy type diagrams correspond to Fig.~\ref{fig2}:
\begin{eqnarray}
	&&K_L(S_3)=\sum_{F=\chi^{\pm},\chi^0}\sum_{S=\tilde{\nu}^{I(R)},\tilde{l}}\frac{m_F m_{l_i}}{m_{l_i}^2-m_{l_j}^2}
	H^{hl_j\bar{l}_j}H_L^{\bar{l}_jFS}H_L^{l_i\bar{F}S^*}I_1(x_F,x_S),\nonumber\\&&K_R(S_3)=K_L(S_3)|_{L{\leftrightarrow}R}.
\end{eqnarray}

The decay widths for processes $h\rightarrow l_i l_j$ are deduced here
\begin{eqnarray}
&&\Gamma(h\rightarrow l_il_j)=(h\rightarrow \bar{l}_il_j)+(h\rightarrow l_i\bar{l}_j),
\end{eqnarray}
where $\Gamma\left(h\rightarrow l_i\bar{l}_j \right)=\frac{1}{16\pi}m_{h}\left(|K_L|^2+|K_R|^2\right)$~\cite{h01,h02}. Correspondingly, the calculations for $h\rightarrow \bar{l}_il_j$ are same as those for $h\rightarrow l_i\bar{l}_j$. Above all, the branching ratios of $h\rightarrow l_il_j$ can be summarized as
\begin{eqnarray}
&&Br(h\rightarrow l_il_j)=\Gamma(h\rightarrow l_il_j)/\Gamma_{h}.
\end{eqnarray}
Here, the total decay width of the SM-like Higgs boson is $\Gamma_{h}\simeq4.1\times10^{-3}$ GeV~\cite{PDG}.

\subsection{Rare decay $h\rightarrow l_i\bar{l}_j$ by MIA method}
In contrast to the mass eigenstate basis, the Feynman diagrams of the MIA assume that external particles are in the mass basis, while internal sparticles remain in the electroweak interaction basis. The diagonal terms of the mass matrix are defined as the sparticle mass, and the off-diagonal terms are defined as the interaction vertices considered as mass insertions~\cite{Dedes2015,Rosiek2016}. In this work, we will compute analytically all the relevant diagrams by applying the MIA in the off-diagonal mass insertions $\Delta^{XX}_{ij}(X=L,R)$, which contribute to the LFV of Higgs boson decays at the one-loop level. The terms $L$ and $R$ specifically denote left-handed sleptons (sneutrinos) and right-handed sleptons.

Concretely, we work with the mass basis for the 
external particles $h$, $l_j$, $l_i$ and with the electroweak interaction basis for the internal 
sparticles in the loops, which from now on will be shortly denoted by: $\tilde{W}^{\pm}$, $\tilde{W}^3$, $\tilde{B}$, $\tilde{B^{'}}$, $\tilde{H}^{\pm}$, $\tilde{H}_{1,2}$, $\tilde{l}_{i}^{L,R}$ ($i = 1, 2, 3$), $\tilde{\nu}_ {L_i}^{I,R}$ ($i = 1, 2, 3$), where $\tilde{l} _{i}^{R}$ is the i-th generation right-handed sleptons, $\tilde{l} _{i}^{L}$ is the i-th generation left-handed sleptons, $\tilde{\nu}^{R,I}_{L_{j}}$ is the j-th generation left-handed CP-even(odd) sneutrino. The LFV comes from the off-diagonal terms of the sleptons (sneutrinos) mass matrix. Thus, we define $(i\ne j)$~\cite{Calibbi2018}
\begin{eqnarray}
	\Delta^{LL}_{ij}=(m_{\tilde{L} }^{2} )_{ij},~~\Delta^{LR}_{ij}=\frac{1}{\sqrt{2} } v_d T_{e,ij},~~\Delta^{RR}_{ij}=(m_{\tilde{E} }^{2} )_{ij}.
\end{eqnarray}
 After our analysis, only these mass insertions change the flavor. For the insertion of $\Delta^{LR}_{ii}$, this case is considered an insertion that does not change flavor. Therefore, it is necessary to introduce an additional insert $\Delta^{LL}_{ij}$ or $\Delta^{RR}_{ij} (i\ne j)$ to change the flavor. 

The LFV processes in the MIA need to consider the trilinear couplings under the interaction eigenstate, so we show some couplings needed in this work as follows. The interactions of lepton-charginos-CP-even(odd) sneutrinos are deduced as:
\begin{eqnarray}
	&&{\cal L}_{\bar{l}_j\chi^-\tilde{\nu}^R}=\frac{i}{\sqrt{2}}\bar{l}_j
	\tilde{\nu}_L^R[Y_e^jP_L\tilde{H}^-+g_2P_R\tilde{W}^-],\nonumber \\&&
	{\cal L}_{\bar{l}_j\chi^-\tilde{\nu}^I}=\frac{1}{\sqrt{2}}\bar{l}_j
	\tilde{\nu}_L^I[-Y_e^jP_L\tilde{H}^-+g_2P_R\tilde{W}^-].
\end{eqnarray}
The lepton-neutralinos-sleptons are deduced as:
\begin{eqnarray}
	&&{\cal L}_{\bar{l}_j\chi^0\tilde{l}}=i\bar{l}_j\Big\{-\Big[\frac{1}{\sqrt{2}}
	\Big(2g_{1}P_{L}\tilde{B}+(g_B+2g_{YB})P_{L}\tilde{B}'\Big)\tilde{l}^R+Y_e^{j}P_{L}\tilde{H}_1\tilde{l}^L\Big]\nonumber \\&&\hspace{2.3cm}+\Big[\frac{1}{\sqrt{2}}
	\Big(g_{2}P_{R}{\tilde{W}^{3}}+g_{1}P_{R}\tilde{B}+(g_B+g_{YB})P_{R}\tilde{B}'\Big)\tilde{l}^L-Y_e^{j}P_{R}\tilde{H}_1\tilde{l}^R\Big]\Big\}.
\end{eqnarray}

\begin{figure}[t]
	\centering
	\includegraphics[scale=0.7]{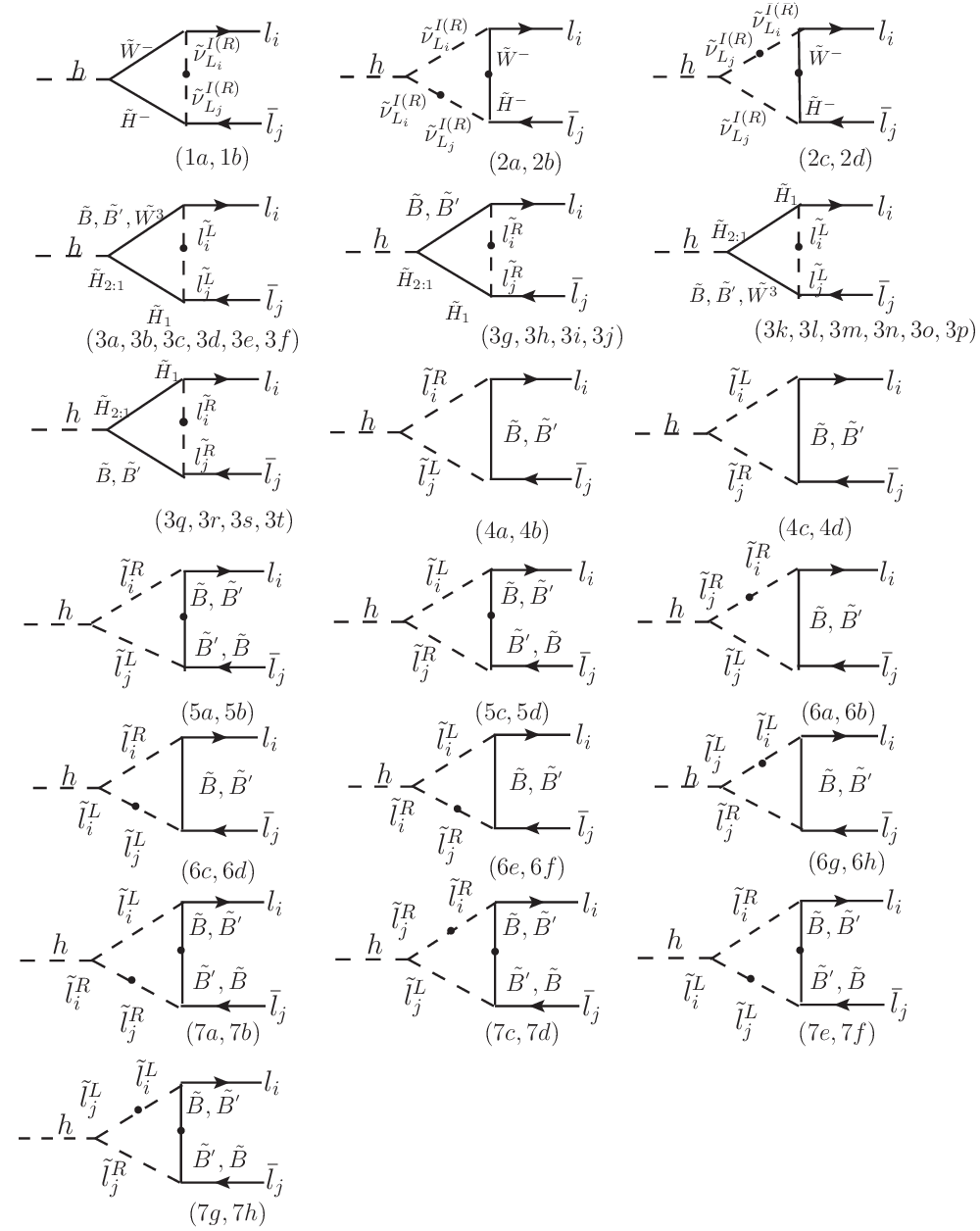}\\
	\caption{The triangle type diagrams for decays $h\rightarrow l_i\bar{l}_j$ in the electroweak interaction basis.} \label{fig3}
\end{figure}

\begin{figure}[t]
	\centering
	\captionsetup{justification=raggedright,singlelinecheck=false}
	\includegraphics[width=13cm]{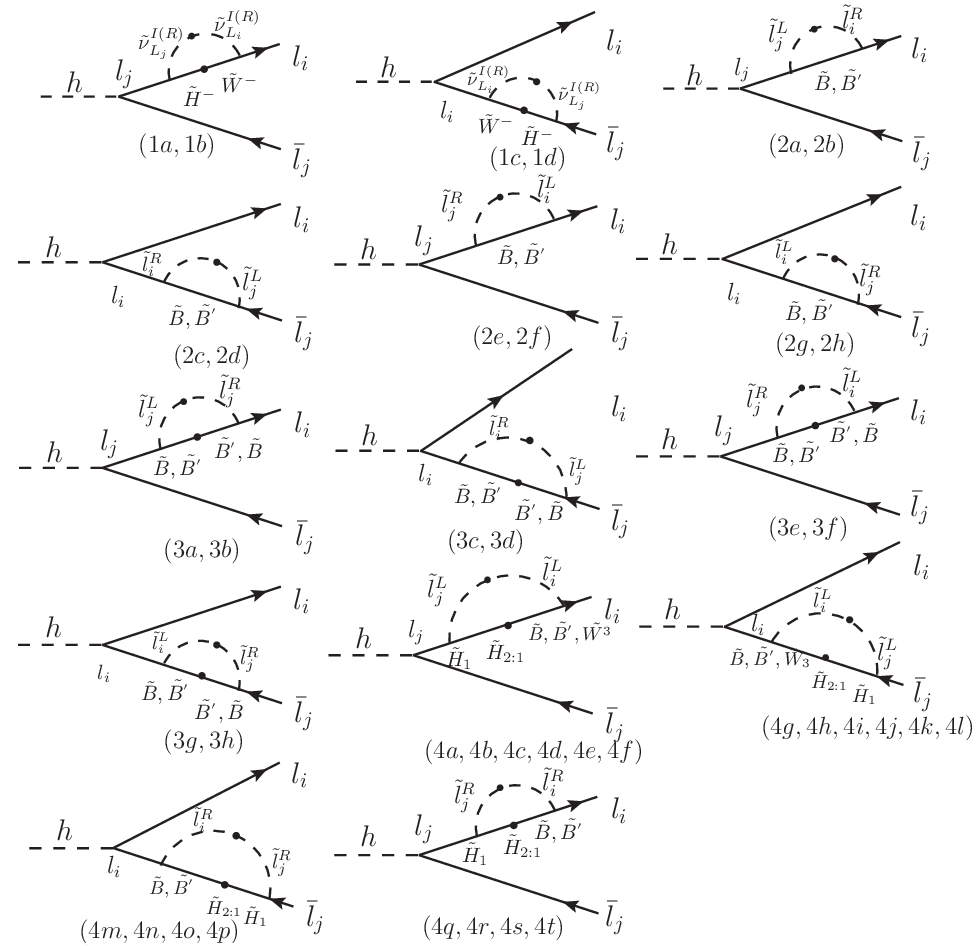}\\
	\caption{The self-energy type diagrams for decays $h\rightarrow l_i\bar{l}_j$ in the electroweak interaction basis.} \label{fig4}
\end{figure}

We show Feynman diagrams using the MIA in Fig.~\ref{fig3} and Fig.~\ref{fig4}. The Feynman diagrams including the right-handed sneutrinos are strongly suppressed by the coupling parameter $Y_{\nu}$, and the situations with right-handed sneutrinos are neglected. To save space, we present the contributions from several Feynman diagrams, the rest of which can be found in the Appendix~\ref{MR}. The one-loop contributions from Fig.~\ref{fig3}(1a) are shown as:
\begin{eqnarray}
	&&K'_{R}(1a)_{\text{MIA}}=-\frac{\sqrt{2}}{4}g_{2}^2 Y_{e,jj}\Delta_{ij}^{LL}[G_{5}(x_{\tilde{\nu}^{I}_{L_{j}}},x_{\tilde{\nu}^{I}_{L_{i}}},x_{2},x_{\mathcal{H} })m_{\mathcal{H}} M_2 Z_{12}^{H} \nonumber\\&&\hspace{1.7cm}+G_{6}(x_{\tilde{\nu}^{I}_{L_{j}}},x_{\tilde{\nu}^{I}_{L_{i}}},x_{2},x_{\mathcal{H} } )  Z_{11}^{H}],\nonumber\\
	&&K'_{L}(1a)_{\text{MIA}}=0.\
	\label{eq38}
\end{eqnarray}
The one-loop contributions from Fig.~\ref{fig3}(3a)
\begin{eqnarray}
	&&K'_{R}(3a)_{\text{MIA}}=\frac{\sqrt{2} }{4}g_{1}^{2} Y_{e,jj}Z_{12}^{H} \Delta_{ij}^{LL}
	G_{5}(x_{\tilde{l}_{j}^{L} },x_{\tilde{l}_{i}^{L} },x_{1},x_{\mathcal{H} } )m_{\mathcal{H} }  M_1,\nonumber\\
	&&K'_{L}(3a)_{\text{MIA}}=0.\
	\label{eq39}
\end{eqnarray}
The one-loop contributions from Fig.~\ref{fig3}(6a)
\begin{eqnarray}
&&K'_{L}(6a)_{\text{MIA}}=g_{1}^{2} P \Delta_{ij}^{RR}G_{5}(x_{\tilde{l}_{j}^{R} },x_{\tilde{l}_{i}^{R} },x_{\tilde{l}_{j}^{L} },x_1) M_1,\nonumber\\
&&K'_{R}(6a)_{\text{MIA}}=0.\
 \label{eq40}
\end{eqnarray}
 Here, $P=\frac{1}{2} (-\sqrt{2} T_{e,jj}Z _{1,1}^{H} +\lambda v_S Y_{e,jj} Z _{12}^{H})$ is coupling vertex, $m_{\mathcal{H}}=\frac{\lambda v_S}{\sqrt{2}}$, \(x_1=\frac{M_{1}^{2}} {\Lambda^2}\), \(x_{\mathcal{H}}=\frac{ m_{\mathcal{H}}^{2}}{\Lambda^2}\), and \(x_{\tilde{l}_{j}^{L} }=x_{\tilde{\nu}^{I,R}_{L_{j}}}=\frac{(M^{2}_{L})^{jj}_{LL}}{\Lambda^2}\). In the equations above, we can derive the parameter $m_{EE}$, $m_{E}$, $m_{LL}$, $m_{L}$, $\delta^{XX}_{ij}(X=L, R)$, which influence the mass matrices of sleptons and sneutrinos thereby affecting the form factors $G_{1}$, $G_{5}$, $G_{6}$, and ultimately impacting the numerical results. The parameters that influence the Higgs mass matrix also affect the numerical results, such as $\kappa$ and $\lambda_{2}$. The parameters $g_{B}$, $\tan\beta$, $\tan\beta'$ and $\lambda$ influence the mass matrices of sleptons, sneutrinos, and Higgs bosons, which in turn affect the decay processes. 
 
 In order to more intuitively analyze the factors that affect LFV processes $h\rightarrow l_i\bar{l}_j$, we suppose that the sparticle masses are degenerate. In other words, we give the one-loop results in the extreme case where the sparticle masses and mass insertion treams are equal to $\Lambda$:
\begin{eqnarray}
	&&M_{1}=M_{2}=m_\mathcal{H}=M_{BL}=M_{BB^{\prime}}=m_L=m_E=m_{LL}=m_{EE}=Ae=\Lambda.
\end{eqnarray}
In this extreme degeneracy case, the corresponding one-loop functions in the electroweak interaction basis are subjected to limited operations, which are equal to some specific values:\begin{eqnarray}
	&&G_{1}(1, 1, 1)=\frac{1}{32\pi^{2}},~~~G_{5}(1, 1, 1, 1)=\frac{1}{96\pi^{2}\Lambda^{4}},~~~G_{6}(1, 1, 1, 1)=-\frac{1}{48\pi^{2}\Lambda^{2}}.
\end{eqnarray}
And mass insertions $\Delta^{XX}_{ij}(X=L, R)$ that change leptons flavor become the product of $\Lambda^{2}$ and the dimensionless parameters $\delta^{XX}_{ij}(X=L, R)$:
\begin{eqnarray}
	&&\Delta^{LL}_{ij}\approx \Lambda^{2} \delta^{LL}_{ij},~~~\Delta^{LR}_{ij}\approx \Lambda^{2} \delta^{LR}_{ij},~~~\Delta^{RR}_{ij}\approx \Lambda^{2} \delta^{RR}_{ij}.
\end{eqnarray}

Eq.~(\ref{eq38})-Eq.~(\ref{eq40}) can be simplified to:
\begin{eqnarray}
	&&K'_{R}(1a)_{\text{MIA}}=-\frac{1}{192\sqrt{2}\pi^{2}}g_{2}^2Y_{e,jj}\delta_{ij}^{LL}Z_{12}^{H}.\nonumber\\
	&&K'_{R}(3a)_{\text{MIA}}=\frac{\sqrt{2} }{4}g_{1}^{2} Y_{e,jj}Z_{12}^{H} \delta_{ij}^{LL}\frac{1}{96\pi^{2}}.\nonumber\\
	&&K'_{L}(6a)_{\text{MIA}}=\frac{1}{96\sqrt{2}\pi^{2}}
	g_{1}^{2}(-Z_{11}^{H}+Y_{e,jj}Z_{12}^{H})  \delta_{ij}^{RR} .
\end{eqnarray}
Then, we use above method to simplify all obtained coefficients to get some approximate results:
\begin{eqnarray}
&&({K'}_{L}^{\delta_{ij}^{LL}}){}_{\text{MIA}}\approx-\frac{\delta_{ij}^{LL}}{192 \sqrt{2}\pi^2}Z_{11}^{H}\Big[2g_1^2+(2g_{YB}+g_B)(g_{YB}+g_B)+g_1(4g_{YB}+3g_B)\Big],\nonumber\\
&&({K'}_{R}^{\delta_{ij}^{LL}}){}_{\text{MIA}}\approx\frac{\delta_{ij}^{LL}}{192 \sqrt{2}\pi^2}\Big[Y_{e,jj}Z_{12}^{H} \Big(-5g_2^2+5g_1^2+(2g_{YB}+g_B)(g_{YB}+g_B)\nonumber\\
&&\hspace{1.2cm}+g_1(4g_{YB}+3g_B)+3g_{YB}(g_{YB}+g_B)\Big) +Z_{11}^{H}\Big((-2+Y_{e,jj}\tan\beta)g_1^2\nonumber\\
&&\hspace{1.2cm}+3Y_{e,jj}\tan\beta g_2^2-(2g_{YB}+g_B)(g_{YB}+g_B)-g_1(4g_{YB}+3g_B)\nonumber\\
&&\hspace{1.2cm}+Y_{e,jj}\tan\beta g_{YB}(g_{YB}+g_B)\Big)\Big],\nonumber\\
&&({K'}_{L}^{\delta_{ij}^{RR}}){}_{\text{MIA}}\approx\frac{\delta_{ij}^{RR}}{192 \sqrt{2}\pi^2}\Big[Y_{e,jj}Z_{12}^{H}\Big(-4g_1^2-3g_{YB}(2g_{YB}+g_B)\nonumber\\
&&\hspace{1.2cm}+(2g_{YB}+g_B)(g_{YB}+g_B)+g_1(4g_{YB}+3g_B)\Big)+Z_{11}^{H}\Big((-2+2Y_{e,jj}\tan\beta)g_1^2\nonumber\\
&&\hspace{1.2cm}-(2g_{YB}+g_B)(g_{YB}+g_B)-g_1(4g_{YB}+3g_B)+g_{YB}(2g_{YB}+g_B)Y_{e,jj}\tan\beta\Big)\Big],\nonumber\\
&&({K'}_{R}^{\delta_{ij}^{RR}}){}_{\text{MIA}}\approx-\frac{\delta_{ij}^{RR}}{192 \sqrt{2}\pi^2}Z_{11}^{H}\Big[2g_1^2+(2g_{YB}+g_B)(g_{YB}+g_B)+g_1(4g_{YB}+3g_B)\Big],\nonumber
\end{eqnarray}
\begin{eqnarray}
&&({K'}_{L,R}^{\delta_{ij}^{LR}}){}_{\text{MIA}}\approx-\frac{\delta_{ij}^{LR}}{32 \sqrt{2}\pi^2}Z_{11}^{H}\Big[g_1^2+\frac{(2g_{YB}+g_B)(g_{YB}+g_B)}{2}\Big]\nonumber\\
&&\hspace{1.2cm}-\frac{\delta_{ij}^{LR}}{96 \sqrt{2}\pi^2}Z_{11}^{H}\Big[\frac{g_1(4g_{YB}+3g_B)}{2}\Big(2+\frac{M_{BL}M_1}{\Lambda^2}\Big)\Big]\nonumber\\
&&\hspace{1.2cm}+\frac{\delta_{ij}^{LR}}{32 \sqrt{2}\pi^2}Z_{11}^{H}\Big[g_1^2+\frac{(2g_{YB}+g_B)(g_{YB}+g_B)}{2}\Big]\nonumber\\
&&\hspace{1.2cm}+\frac{\delta_{ij}^{LR}}{96 \sqrt{2}\pi^2}Z_{11}^{H}\Big[\frac{g_1(4g_{YB}+3g_B)}{2}\Big(\frac{2M_{BB'}}{\Lambda}+\frac{M_{BL}M_1M_{BB'}}{\Lambda^3}\Big)\Big].
\label{45}
\end{eqnarray}

When the sparticle masses are degenerate, we can clearly see that $\delta^{XX}_{ij}(X=L, R)$ are sensitive parameters and have an important impact. Additionally, the contribution of $\delta^{LR}_{ij}$ is close to zero and can be ignored. With the supposition $g_{YB}+g_B>0, 2g_{YB}+g_B<0, g_{YB}<0, g_B>0, |3g_{YB}(2g_{YB}+g_B)|>|g_1(4g_{YB}+3g_B)|>|(2g_{YB}+g_B)(g_{YB}+g_B)|, |3g_{YB}(g_{YB}+g_B)|>|g_1(4g_{YB}+3g_B)|, g_1 \approx 0.3, g_2 \approx 0.6, Y_{e,jj}\tan\beta\approx0.4(j=2)$. The order analysis shows 
\begin{eqnarray}
	&&0<\left(\frac{(2g_{YB}+g_B)(g_{YB}+g_B)+g_1(4g_{YB}+3g_B)}{2g_1^2}\right)<1, \nonumber\\
    &&0<\left(\frac{-3g_{YB}(2g_{YB}+g_B)+(2g_{YB}+g_B)(g_{YB}+g_B)+g_1(4g_{YB}+3g_B)}{-4g_1^2}\right)<1,\nonumber\\
    &&0<\left(\frac{(2g_{YB}+g_B)(g_{YB}+g_B)+g_1(4g_{YB}+3g_B)+3g_{YB}(g_{YB}+g_B)}{-5g_2^2}\right)<1. \
\end{eqnarray}
The analysis above clearly demonstrates that parameters $g_{YB}$ and $g_B$, as new couplings, produce a positive correction to the LFV. 
\section{Numerical Results}

In this section, we analyze the numerical results and consider the experimental constraints. Given that experimental constraints from the $l_j\rightarrow l_i \gamma$ processes tightly constrain the parameter space of the NB-LSSM, it is crucial to take into account the impacts of $l_j\rightarrow l_i \gamma$ when analyzing the processes of $h\rightarrow l_il_j$. We study some sensitive parameters on these processes, which are detailed below. To take the numerical calculation, some restrictions should be taken into account.
\begin{enumerate}
	\item Considering the updated experimental data on searching \( Z' \) indicates \( M_{Z'} \geq 5.15 \, \text{TeV} \) at 95\% C.L.~\cite{CMS:2022eud}, we choose \( M_{Z'} = 5.15 \, \text{TeV} \) in the follows.
	\item The ratio between the \( Z' \) mass and its gauge coupling at 99\% C.L. as \( M_{Z'}/g_B \geq 6 \, \text{TeV} \)~\cite{GCG,MAB}, and then the scope of \( g_B \) is \( 0 < g_B < 0.86 \).
	\item LHC experimental data constrain \( \tan \beta' < 1.5 \)~\cite{48}.
	\item For particles that exceed the SM, the mass limits considered are: the slepton mass is greater than 700 GeV, the neutralino mass is larger than 500 GeV, the chargino mass is greater than 900 GeV~\cite{navas2024} and the charged Higgs mass is greater than 600 GeV~\cite{ATLAS:2024hya}.
	\item The lightest CP-even Higgs mass is around \( m_h = 125.20 \pm 0.11 \, \text{GeV} \)~\cite{ATLAS:2023oaq}.
	\item The Higgs $h$ decays$(h \rightarrow \gamma\gamma, ZZ^*, WW^*, b \bar b, \tau \bar\tau)$ should be satisfied~\cite{ATLAS:LFVHiggs2023}.
\end{enumerate}

The relevant SM input parameters are chosen as \( m_W = 80.385 \, \text{GeV} \), \( m_Z = 91.188 \,\text{GeV} \), \( \alpha_{\text{em}}(m_Z) = \frac{1}{128.9} \), \( \alpha_s(m_Z) = 0.119 \). Considering the constraint of the experiments~\cite{navas2024}, we take \( M_1 = 700 \, \text{GeV} \), \( M_2 = 600 \, \text{GeV} \), \( M_{BB^{'} } = 800 \, \text{GeV} \),  \( M_{BL} = 900 \, \text{GeV} \), \( m_{\tilde{q}} = m_{\tilde{u}} = \text{diag}(2, 2, 2) \, \text{TeV} \), and \( T_u = \text{diag}(1, 1, 1) \, \text{TeV} \), respectively.
\begin{figure}[!htbp]
	\centering
	\captionsetup{justification=raggedright,singlelinecheck=false}
	\newcommand{\subfigwidth}{0.32\linewidth}
	\newcommand{\subfigheight}{5cm}
	\begin{subfigure}[t]{\subfigwidth}
		\includegraphics[width=\linewidth,height=\subfigheight,keepaspectratio=false]{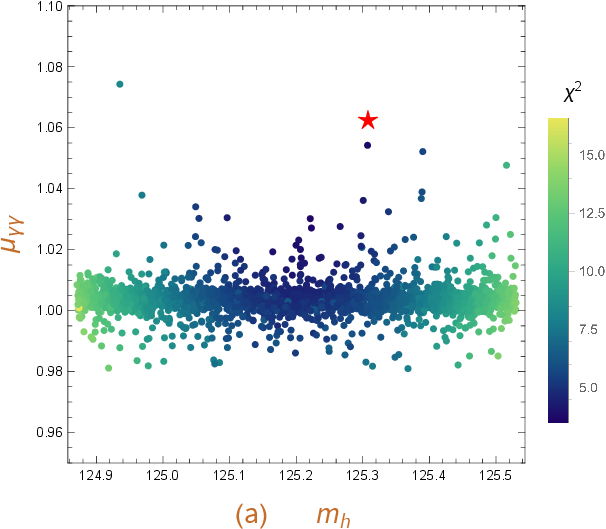}
		\caption*{} 
		\label{fig:sub1}
	\end{subfigure}\vspace{-0.8cm}
	\hfill
	\begin{subfigure}[t]{\subfigwidth}
		\includegraphics[width=\linewidth,height=\subfigheight,keepaspectratio=false]{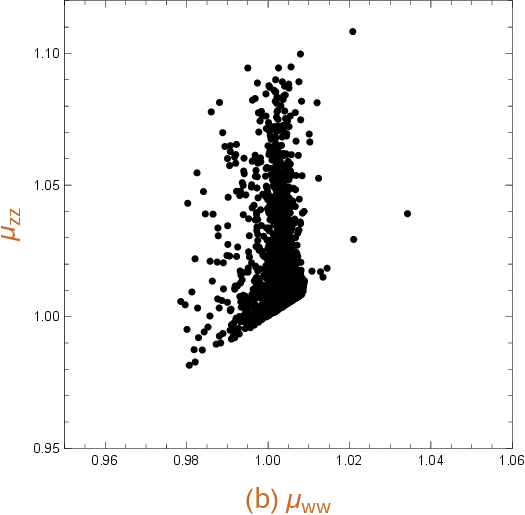}
		\caption*{}
		\label{fig:sub2}
	\end{subfigure}
	\hfill
	\begin{subfigure}[t]{\subfigwidth}
		\includegraphics[width=\linewidth,height=\subfigheight,keepaspectratio=false]{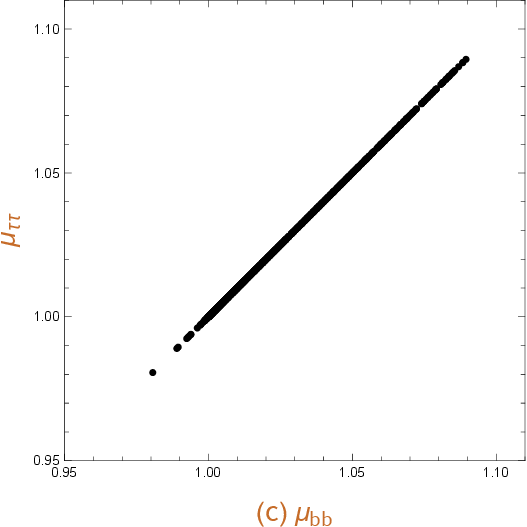}
		\caption*{}
		\label{fig:sub3}
	\end{subfigure}
	\vspace{-1cm}
	\begin{subfigure}[t]{\subfigwidth}
		\includegraphics[width=\linewidth,height=\subfigheight,keepaspectratio=false]{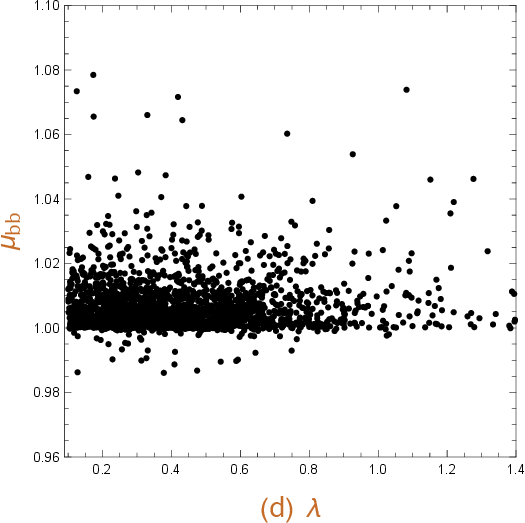}
		\label{fig:sub4}
	\end{subfigure}
	\hspace{0.05\linewidth}
	\begin{subfigure}[t]{\subfigwidth}
		\includegraphics[width=\linewidth,height=\subfigheight,keepaspectratio=false]{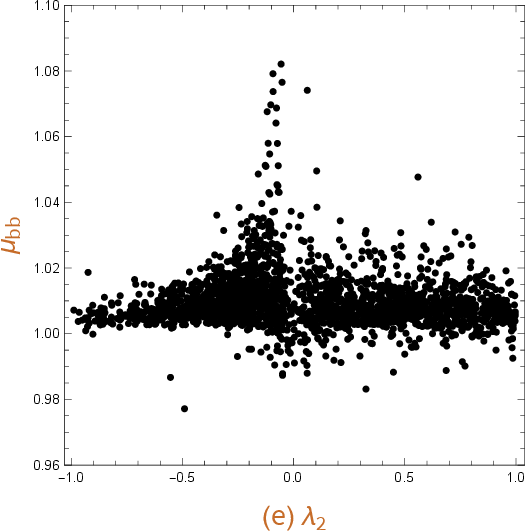}
		\label{fig:sub5}
	\end{subfigure}
	\vspace{0.1cm}
	\caption{The scatter points for Higgs mass and signal strengths. The red star represents the best-fit point ($\chi^2_{\mathrm{best}} = 3.480$). All scattered points satisfy the $3\sigma$ experimental constraints of Higgs mass and the $2\sigma$ experimental constraints of corresponding Higgs decays.}
	\label{fig:combined}
\end{figure}

 Firstly, the 125 GeV Higgs boson mass and the Higgs signal strengths 
	$\mu_{\gamma\gamma, WW^*, ZZ^*, b\bar{b}, \tau\bar{\tau}}$ have been incorporated into our analysis. Under the premise of satisfying the above experimental constraints, some free parameters are selected for random scanning as follows: $\tan\beta \in (5,60)$, $\tan\beta' \in (1,1.5)$, $g_B \in (0.1,0.85)$, $g_{YB} \in (-0.45,-0.05)$, $v_S \in (0.5,5)\,\text{TeV}$, $\kappa \in (-1,3)$, $\lambda \in (1,1.4)$, $\lambda_2 \in (-10,10)$, $T_\kappa \in (-2,3)\,\text{TeV}$, $T_\lambda \in (-2,2)\,\text{TeV}$, $T_{\lambda_2} \in (-2,2)\,\text{TeV}$. In this work, we perform a $\chi^2$ test to explore the best-fit points describing 125 GeV Higgs mass and the signal strength in the model. Generally, the $\chi^2$ function can be constructed as
	\begin{equation}
		\chi^2 = \sum_{i=1}^n \left( \frac{\mu_i^{\text{th}} - \mu_i^{\text{exp}}}{\delta _i^{\text{exp}}} \right)^2,
		\label{eq:chi2}
\end{equation}
where \(\mu_i^{\text{th}}\) denotes the theoretical prediction for the \(i\)th observable, \(\mu_i^{\text{exp}}\) is the corresponding experimental measurement, and \(\delta _i^{\text{exp}}\) represents the experimental uncertainty. The averaged values of the experimental data are derived from the updated PDG, $m^{exp}_{h}$ = 125.20 $\pm$ 0.11 GeV, $\mu^{exp}_{\gamma\gamma}$ = 1.10 $\pm$ 0.06, $\mu^{exp}_{ZZ^*}$ = 1.02 $\pm$ 0.08, $\mu^{exp}_{WW^*}$ = 1.0 $\pm$ 0.08, $\mu^{exp}_{b\bar b}$ = 0.99 $\pm$ 0.12, $\mu^{exp}_{\tau\bar\tau}$ = 0.91 $\pm$ 0.09~\cite{cms2022,atlas-cms2016,cdf-d02013,atlas2020,atlas2021a,atlas2021b,atlas2019}. The parameter set that minimizes this \(\chi^2\) function constitutes the best-fit solution. The reasonable parameter space is selected for scattered points, as shown in Fig.~\ref{fig:combined}.

In Fig.~\ref{fig:combined}(a), we present the parameter space scan results of the Higgs mass versus the signal strength of $\mu_{\gamma\gamma}$. One can clearly observe that as the $\chi^2$ value decreases, the scattered points of $m_h$ gradually converge to the central value of the experimental measurement. Through computational verification, the signal strength $\mu_{\gamma\gamma}$, $\mu_{WW^*}$, $\mu_{ZZ^*}$, $\mu_{\tau\tau}$ and $\mu_{b\bar{b}}$ all satisfy the 95\% confidence level constraint ($\chi^2 < 19.68$ with 11 freedom parameters). Additional, the Fig.~\ref{fig:combined}(b) and (c) show that the values of $\mu_{ZZ^*}$ ($\mu_{\tau\bar{\tau}}$) overall increase with increasing value of $\mu_{WW^*}$ ($\mu_{b\bar{b}}$), which indicate that the four channels of $h \to ZZ^*, WW^*, b\bar{b}, \tau\bar{\tau}$ possess the similar characters. Then signal strength $\mu_{b\bar{b}}$ as a function of the viable parameters $\lambda$ and $\lambda_2$ are figured out in Fig.~\ref{fig:combined}(d) and (e), constrained by Higgs mass and signal strength measurements. We can see that the value range of $\lambda$ is around 0.1-0.8, where $\mu_{b\bar{b}}$ shows moderate variation while remaining compatible with observations. In Fig.~\ref{fig:combined}(e), the  $\mu_{b\bar{b}}$ depends on $\lambda_2$, where $\lambda_2$ is restricted to the interval $[-1, 1]$. For $\lambda_2 < 0$, $\mu_{b\bar{b}}$ exhibits slight enhancement as $|\lambda_2|$ decreases, while the signal strength remains stable for $\lambda_2 > 0$. Other parameters can make Higgs mass and decays find suitable signals within the range of the scan. This demonstrates the viability of the chosen parameter space in reproducing the observed data. From the Fig.~\ref{fig:combined}(a)-(e), we also observe that the Higgs signals in the NB-LSSM framework exhibit $\sim 10\%$ new physics corrections compared to the SM predictions, providing a significant guidance for new physics searches.
\begin{figure}[!htbp]
	\centering
	\captionsetup{justification=raggedright,singlelinecheck=false}
	\newcommand{\subfigwidth}{0.48\linewidth}
	\newcommand{\subfigheight}{5cm}
	
	\begin{subfigure}[b]{\subfigwidth}
		\includegraphics[width=\linewidth,height=\subfigheight,keepaspectratio]{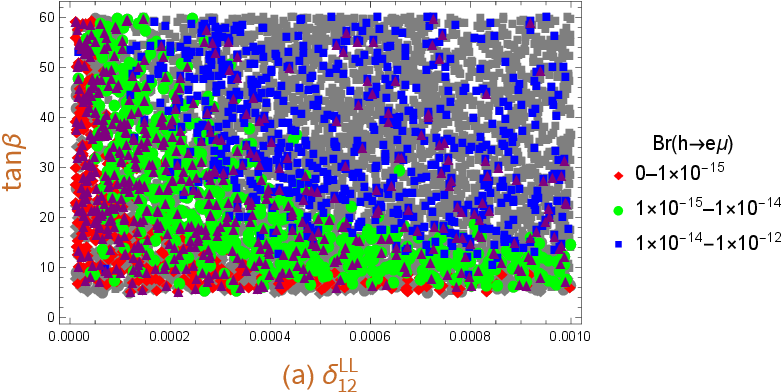}
		\caption*{}
		\label{fig:sub6}
	\end{subfigure}\vspace{-1cm}
	\hfill
	\begin{subfigure}[b]{\subfigwidth}
		\includegraphics[width=\linewidth,height=\subfigheight,keepaspectratio]{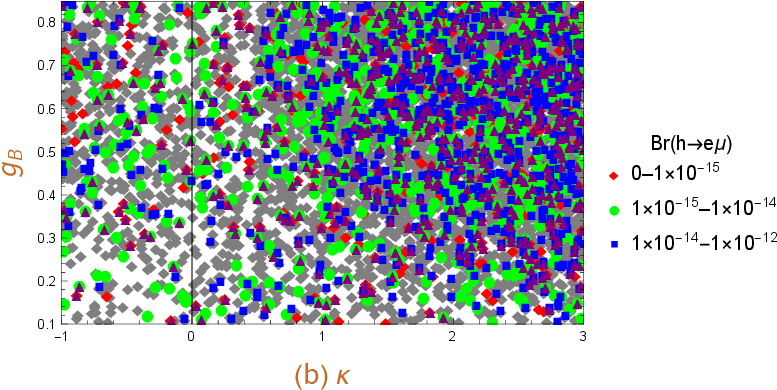}
		\caption*{}
		\label{fig:sub7}
	\end{subfigure}
	\vspace{-1cm}
	\hspace*{-1cm}
	\begin{subfigure}[b]{\subfigwidth}
		\includegraphics[width=\linewidth,height=4cm,keepaspectratio]{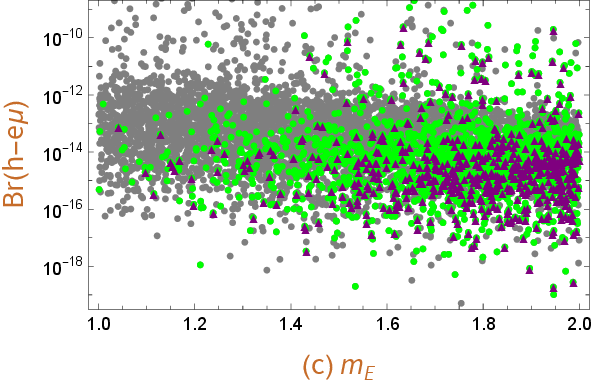}
		\caption*{}
		\label{fig:sub8}
	\end{subfigure}
	\hfill
	\begin{subfigure}[b]{\subfigwidth}
		\includegraphics[width=\linewidth,height=\subfigheight,keepaspectratio]{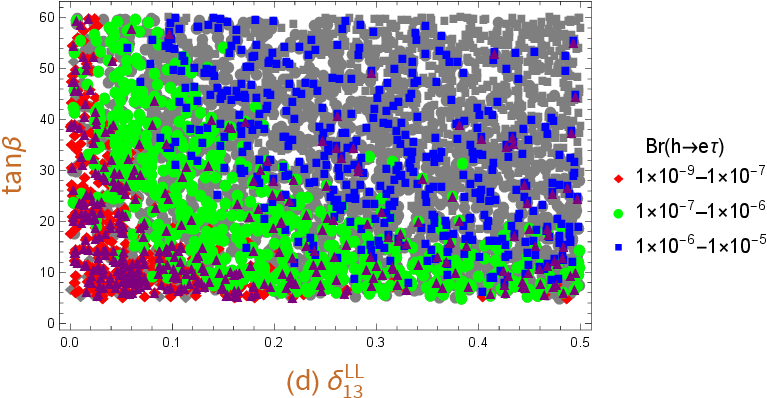}
		\caption*{}
		\label{fig:sub9}
	\end{subfigure}
	\vspace{-1cm}
	\centering
	\begin{subfigure}[b]{\subfigwidth}
		\includegraphics[width=\linewidth,height=\subfigheight,keepaspectratio]{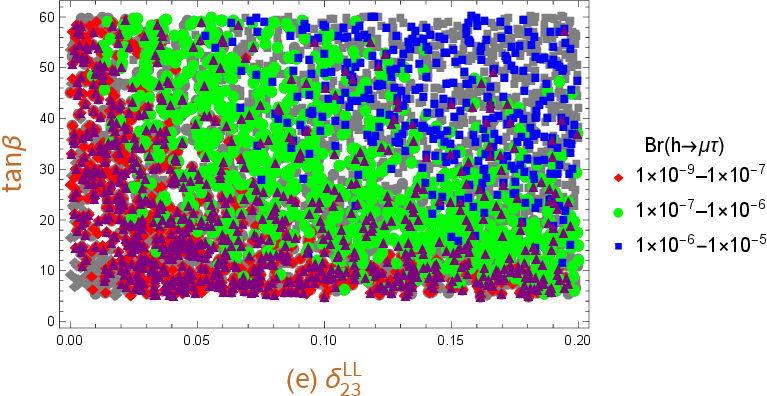}
		\caption*{}
		\label{fig:sub10}
	\end{subfigure}
	\vspace{0.1cm}
	\caption{Parameter space scans for LFV processes $h \to e\mu, e\tau$ and $\mu\tau$. Gray regions denote the parameter space allowed by the Higgs mass within $3\sigma$ experimental limits and the signal strength within $2\sigma$ experimental limits. The colored ($\textcolor{red}{\blacklozenge}$, $\textcolor{green}{\bullet}$ and $\textcolor{blue}{\blacksquare}$) regions incorporate additional LFV constraints of $\text{Br}(l_j \to l_i\gamma)$. $\textcolor{purple}{\blacktriangle}$ denotes the projected future experimental sensitivities for the $\mu \to e\gamma$ and $\tau \to e\gamma\,(\mu\gamma)$ processes. Fig.~\ref{fig:combined_plots}(a)-(c): Dependence of $\text{Br}(h \to e\mu)$ on $\tan\beta$, $\kappa$, $g_B$, $\delta_{12}^{\text{LL}}$, and $m_E$. Fig.~\ref{fig:combined_plots}(d) and (e): Correlations of $\text{Br}(h \to e\tau)$ and $\text{Br}(h \to \mu\tau)$ with $\tan\beta$, $\delta_{13}^{\text{LL}}$, and $\delta_{23}^{\text{LL}}$.}
	\label{fig:combined_plots}
\end{figure}
\begin{table}[!h]
	{ 
		\centering
		\setlength{\tabcolsep}{20pt}
		\caption{Scanning parameters for Fig.~\ref{fig:combined_plots}} 
	    \label{tab:scan_params}}
		\begin{tabular}{|l|c|c|}
			\hline
			\textbf{Parameters}       & \textbf{Min}       & \textbf{Max}       \\ \hline
			\( m_E / \text{TeV} \)   & \( 1 \) & \( 2 \) \\ \hline
			\( v_S / \text{TeV} \)      & 0.5               & 5              \\ \hline
			\( T_{2}, T_{\lambda} / \text{TeV} \)  & -2                  & 2                \\ \hline
			\( T_{\kappa} / \text{TeV} \)  & -2                  & 3                \\ \hline
			\(\tan \beta\)             & 5                  & 60                 \\ \hline
			\(\tan \beta'\)             & 1                  & 1.5                 \\ \hline
			\( g_B \)                  & 0.1                & 0.85                \\ \hline
			\( g_{YB} \)               & 0.05               & 0.45                \\ \hline
			\(\lambda\)               & 0.1                & 1.4                \\ \hline
			\(\lambda_2\)               & -1                & 1                \\ \hline
			\(\kappa\)               & -1                & 3                \\ \hline
			\( \delta _{12}^{LL}, \delta _{12}^{RR}, \delta _{12}^{LR} \)   &  \( 1 \times 10^{-5} \) & \( 1 \times 10^{-3} \) \\ \hline
			\( \delta _{13}^{LL}, \delta _{13}^{RR}, \delta _{13}^{LR} \)   &  \( 1 \times 10^{-3} \) & \( 5 \times 10^{-1} \) \\ \hline
			
			\( \delta _{23}^{LL}, \delta _{23}^{RR}, \delta _{23}^{LR} \)   &  \( 1 \times 10^{-3} \) & \( 2 \times 10^{-1} \) \\ \hline
	\end{tabular}
\end{table}

Then, taking into account the constraints from the Higgs boson mass, signal strengths, and LFV, we perform a comprehensive parameter scan over all relevant benchmark scenarios to explore the allowed parameter space. In conclusion to obtain a more transparent numerical results, we adopt the following assumptions on parameter space: $Ae=0.4\mathrm{TeV},~m_{EE}=m_{LL}=1.5 \mathrm{TeV}, \Lambda=1\mathrm{TeV}$. The reasonable parameter space is selected to scatter points, which are shown in Table~\ref{tab:scan_params}.

In Figs.~\ref{fig:combined_plots}(a)-(e), the gray regions represent the parameter space constrained solely by Higgs mass and signal strength measurements, while the colored ($\textcolor{red}{\blacklozenge}$, $\textcolor{green}{\bullet}$ and $\textcolor{blue}{\blacksquare}$) regions incorporate additional LFV constraints. The comparison clearly demonstrates that the inclusion of present LFV observables leads to a significant reduction in allowed parameter points. $\textcolor{purple}{\blacktriangle}$ denotes the projected future experimental sensitivities for different processes: for the $\mu \to e\gamma$  process, the future sensitivity is expected to reach $\text{Br}(\mu \to e\gamma) \sim 6 \times 10^{-14}$~\cite{Baldini2013}, while for the $\tau \to e\gamma(\mu\gamma)$ processes, the future experimental sensitivities will improve to $\text{Br}(\tau \to e\gamma(\mu\gamma)) \sim 10^{-8}$-$10^{-9}$~\cite{Hayasaka2013}, these future constraints will significantly reduce the allowed parameter space, with substantial portions becoming testable in next-generation experiments.

Specifically, Fig.~\ref{fig:combined_plots}(a) and (b) show the dependence of $\text{Br}(h \to e\mu)$ on the parameters $\delta_{12}^{\text{LL}}$, $\tan\beta$, $\kappa$ and $g_B$. $\textcolor{red}{\blacklozenge}$ represents the  $\text{Br}(h \to e\mu) < 1 \times 10^{-15}$, $\textcolor{green}{\bullet}$ represents the $\text{Br}(h \to e\mu)$ in the range of $1 \times 10^{-15} < \text{Br}(h \to e\mu) < 1 \times 10^{-14}$ and $\textcolor{blue}{\blacksquare}$ represents the $\text{Br}(h \to e\mu)$ in the range of $1 \times 10^{-14} < \text{Br}(h \to e\mu) < 1 \times 10^{-12}$. The $\text{Br}(h \to e\mu)$ shows a clear positive correlation with both the flavor mixing parameter $\delta_{12}^{\text{LL}}$ and $\tan\beta$, exhibiting a monotonic increase from $\mathcal{O}(10^{-15})$ to $\mathcal{O}(10^{-12})$ as these parameters grow. This parameter dependence demonstrates how increasing $\delta_{12}^{\text{LL}}$ and $\tan\beta$ cooperatively enhance the $\text{Br}(h \to e\mu)$ through combined effects in the slepton and Higgs sectors. Although the specific impact of $\kappa$ and $g_B$ on the $\text{Br}(h \to e\mu)$ cannot be clearly discerned, we can observe that their significant influence on the viable parameter space is evident. Their scattered points mostly fall in the upper-right region, indicating that $\kappa$ takes values in the range of 0.5–3, and the scanned parameter space of $g_B$ remains reasonably consistent with experimental constraints. This suggests that these parameters are among the more sensitive ones in the analysis. Fig.~\ref{fig:combined_plots}(c) shows the correlation between the parameter $m_E$ and the $\text{Br}(h \to e\mu)$. It is evident that the $\mu \to e\gamma$ constraints significantly restrict the allowed parameter space. The increasing number of parameter points satisfying experimental constraints with larger $m_E$ values indicates that the diagonal elements of the slepton (sneutrino) mass matrices significantly influence $\text{Br}(h \to e\mu)$. 

Meanwhile, Fig.~\ref{fig:combined_plots}(d) and (e) display the relationships between the parameters $\delta_{13}^{\text{LL}}$, $\delta_{23}^{\text{LL}}$ and $\tan\beta$ with the $\text{Br}(h \to e\tau)$ and $\text{Br}(h \to \mu\tau)$, respectively. $\textcolor{red}{\blacklozenge}$ represents the  $\text{Br}(h \to e\tau(\mu\tau))$ in the range of $1 \times 10^{-9} < \text{Br}(h \to e\tau(\mu\tau)) < 1 \times 10^{-7}$, $\textcolor{green}{\bullet}$ represents the $\text{Br}(h \to e\tau(\mu\tau))$ in the range of $1 \times 10^{-7} < \text{Br}(h \to e\tau(\mu\tau)) < 1 \times 10^{-6}$ and $\textcolor{blue}{\blacksquare}$ represents the $\text{Br}(h \to e\tau(\mu\tau))$ in the range of $1 \times 10^{-6} < \text{Br}(h \to e\tau(\mu\tau)) < 1 \times 10^{-5}$. A clear trend is observed: as the parameters $\delta_{13}^{\text{LL}}$, $\delta_{23}^{\text{LL}}$ and $\tan\beta$ increase, both $\text{Br}(h \to e\tau)$ and $\text{Br}(h \to \mu\tau)$ can be enhanced from $\mathcal{O}(10^{-9})$ to $\mathcal{O}(10^{-5})$. As a result, these parameters play a beneficial role in our analysis.
 
\subsection{125 GeV Higgs LFV decay $h\rightarrow e\mu $ }

\begin{figure}
   	\setlength{\unitlength}{1mm}
   	\centering
   	\captionsetup{justification=raggedright,singlelinecheck=false}
   	\includegraphics[width=2.8in]{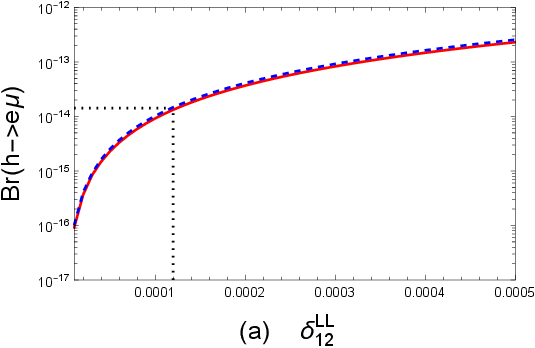}
   	\vspace{0cm}
   	\includegraphics[width=2.8in]{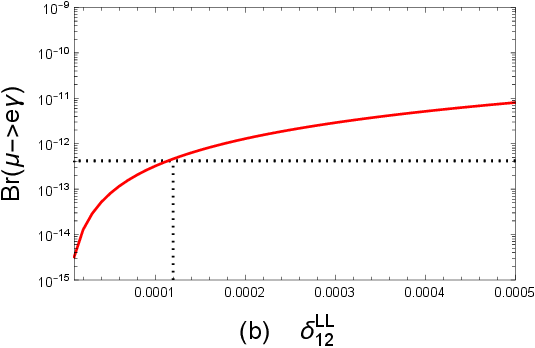}
   	\vspace{0cm}
   	\includegraphics[width=2.8in]{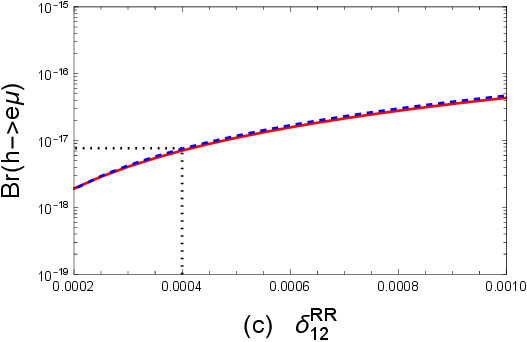}
   	\vspace{0cm}
   	\includegraphics[width=2.8in]{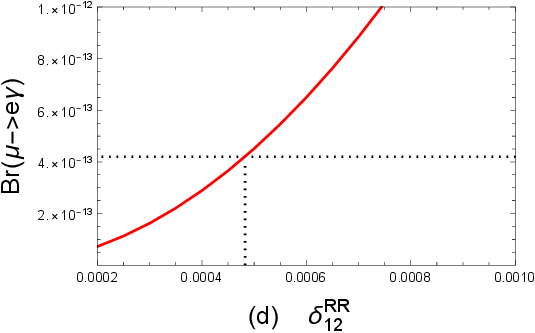}
   	\vspace{0cm}
   	\includegraphics[width=2.8in]{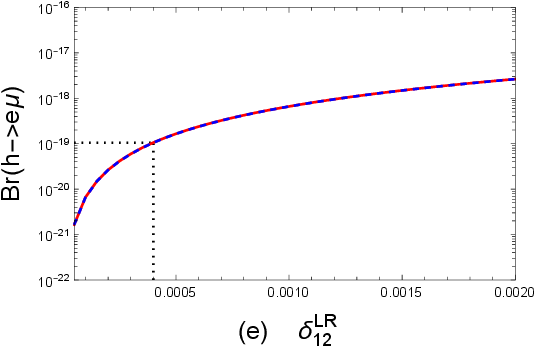}
   	\vspace{0cm}
   	\includegraphics[width=2.8in]{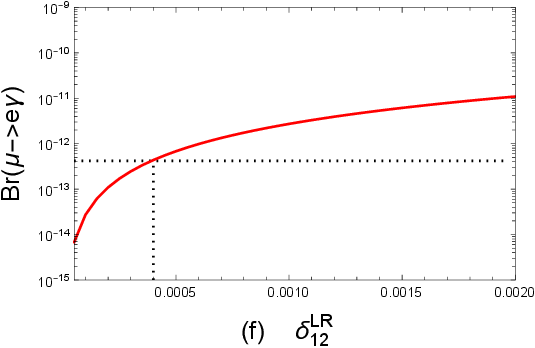}
   	\vspace{0cm}
   	\caption[]{{\label{fig6}} ${\rm{Br}}(h\rightarrow e\mu)$ and ${\rm{Br}}(\mu\rightarrow e\gamma)$ versus slepton flavor mixing parameters $\delta_{12}^{LL}$, $\delta_{12}^{RR}$ and $\delta_{12}^{LR}$, where the red dashed line represents the results in the mass eigenstate basis, the blue dashed line represents the results in the electroweak interaction basis, the black dotted line denotes the present experimental limit of ${\rm{Br}}(\mu\rightarrow e\gamma)$, the represents of the lines follows are the same as here.}
   \end{figure}

  To clarify the parameter dependence of LFV observables, we perform a detailed one-dimensional curve graph analysis under the following experimental constraints: the Higgs mass within its $3\sigma$ experimental limits, Higgs decays complying with $2\sigma$ limits, and LFV processes satisfying both current limits and projected future sensitivities. First, we analyze the exclusive contribution of $\delta^{XX}_{12}$ by nullifying $\delta^{XX}_{13}$ and $\delta^{XX}_{23}$ (X = L, R), and then plot the influence of slepton flavor mixing parameters $\delta^{XX}_{12}$ on the branching ratios of $h\rightarrow e\mu$ and $\mu\rightarrow e\gamma$ in Fig.~\ref{fig6}. In the figures, the latest experimental upper limits of the ${\rm{Br}}(\mu\rightarrow e\gamma)$ are displayed as the black dotted lines.  
By comparing the calculation results of the curves in the two different eigenstates, we can see that the variations in the curves of parameters $\delta^{LL}_{12}$, $\delta^{RR}_{12}$ and $\delta^{LR}_{12}$ are essentially consistent. It can be concluded the accuracy of the MIA results is verified. We can clearly see that when the slepton mixing parameters $\delta_{12}^{XX}$ $(X=L, R)$ increase, both ${\rm{Br}}(h\rightarrow e\mu)$ and ${\rm{Br}}(\mu\rightarrow e\gamma)$ grow rapidly, which means that ${\rm{Br}}(h\rightarrow e\mu)$ and ${\rm{Br}}(\mu\rightarrow e\gamma)$ can achieve significantly enhance within a very small parameter space of $\delta^{XX}_{12}$. Therefore, we deduce that parameters $\delta^{XX}_{12}(X=L, R)$ are sensitive parameters and have a strong effect on the LFV. The reason is that LFV processes are flavor dependent, and LFV rate for $ \mu\rightarrow e\gamma$ and $ h\rightarrow e\mu$ depends on the slepton mixing parameters $\delta_{12}^{XX}~(X=L,R)$. Additionally, it can be seen that ${\rm{Br}}(\mu\rightarrow e\gamma)$ exceeds the experimental limit quickly, which leads to $\delta^{LL}_{12}\lesssim0.00012$, $\delta^{RR}_{12}\lesssim0.0004$, $\delta^{LR}_{12}\lesssim0.0004$. Consequently, the stringent experimental limit on ${\rm{Br}}(\mu\rightarrow e\gamma)$ imposes strong constraints on ${\rm{Br}}(h\rightarrow e\mu)$, rendering the latter difficult to approach its current experimental upper limit. Under these constraints, ${\rm{Br}}(h\rightarrow e\mu)$ can only reach up to $\mathcal{O}(10^{-14})$. 
  \begin{figure}
 	\setlength{\unitlength}{1mm}
 	\centering
 	\includegraphics[width=3in]{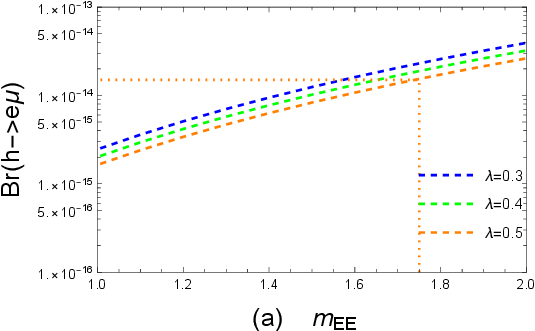}\includegraphics[width=3in]{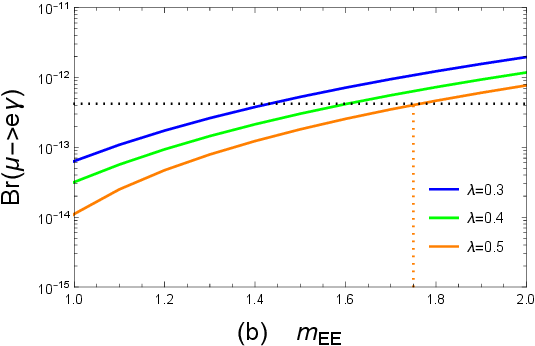}\\
 	
 	\includegraphics[width=3in]{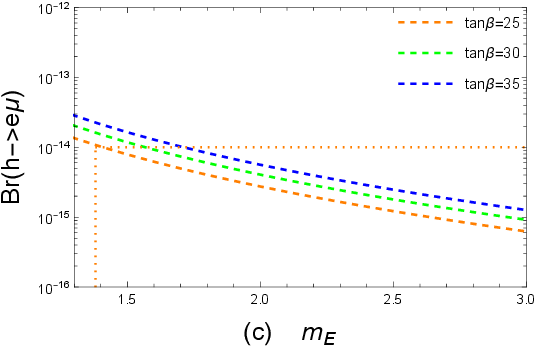}\includegraphics[width=3in]{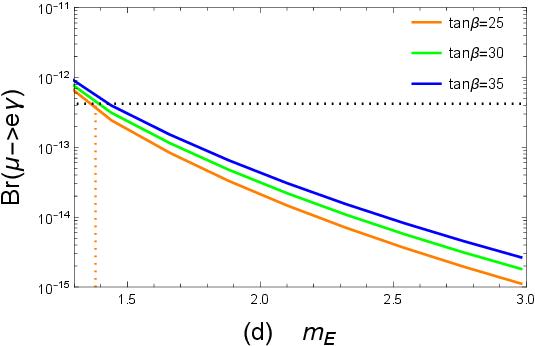}\\
 
 	\includegraphics[width=3in]{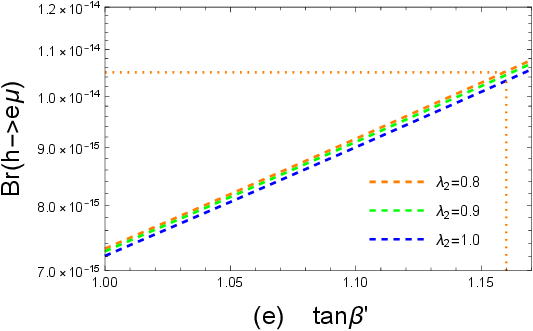}\includegraphics[width=3in]{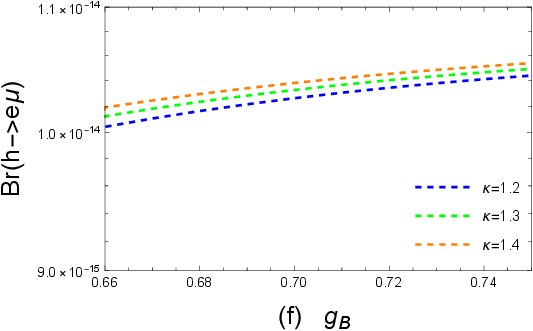}\\
 	
 	\caption[]{{\label{fig7}}${\rm{Br}}(h\rightarrow e\mu)$ and ${\rm{Br}}(\mu\rightarrow e\gamma)$ versus basic parameters $m_{EE}, m_{E}, \tan\beta', g_B$.}
 \end{figure}  
 
 Additionally, we investigate the impact of the basic parameters on ${\rm{Br}}(h\rightarrow e\mu)$ and ${\rm{Br}}(\mu\rightarrow e\gamma)$. For this analysis, we fix the slepton flavor mixing parameters at $\delta_{12}^{LL}=\delta_{12}^{RR}=\delta_{12}^{LR}=1\times 10^{-4}$ to isolate their effects. Then we research the influence of the basic parameters $m_{EE}=m_{LL}$, $m_{E}=m_{L}$, $\tan\beta$, $\tan\beta'$, $\lambda$, $\lambda_{2}$, $\kappa$, and $g_{B}$ on ${\rm{Br}}(h\rightarrow e\mu)$ and ${\rm{Br}}( \mu\rightarrow e\gamma)$, respectively, which can be intuitively seen in Fig.~\ref{fig7}.
 
 The dotted line in Fig.~\ref{fig7} still represents the present limit of ${\rm{Br}}(\mu\rightarrow e\gamma)$. With $\lambda = 0.3, 0.4, 0.5$, we show ${\rm{Br}}(h\rightarrow e\mu)$ and ${\rm{Br}}(\mu\rightarrow e\gamma)$  versus $m_{EE}$ in Fig.~\ref{fig7}(a) and Fig.~\ref{fig7}(b). With $\tan\beta = 25, 30, 35$, the ${\rm{Br}}(h\rightarrow e\mu)$ and ${\rm{Br}}(\mu\rightarrow e\gamma)$ versus the parameter $m_{E}$ are plotted in Fig.~\ref{fig7}(c) and Fig.~\ref{fig7}(d). The parameters $m_{E}$ and $m_{EE}$ located in the diagonal and non-diagonal elements of slepton (sneutrino) matrices have a large effect on the LFV. It can be seen that the parameters $m_{E}$ and $m_{EE}$ are constrained by the present limit $m_{EE}\lesssim1.75{\rm TeV}$ and $m_{E}\gtrsim1.3{\rm TeV}$ with $\lambda = 0.5$ and $\tan\beta = 25$. Notably, both ${\rm{Br}}(h\rightarrow e\mu)$ and ${\rm{Br}}(\mu\rightarrow e\gamma)$ exhibit distinct dependencies on the slepton mass parameters. Besides, it is obvious that ${\rm{Br}}(h\rightarrow e\mu)$ and ${\rm{Br}}(\mu\rightarrow e\gamma)$ increase quickly with the increasing $m_{EE}$, hence $m_{EE}$ has a positive effect on LFV. However, the ${\rm{Br}}(h\rightarrow e\mu)$ and ${\rm{Br}}(\mu\rightarrow e\gamma)$ decrease quickly with the increasing $m_{E}$, the mass of sleptons increases as $m_{E}$ increases, which indicates that heavy sleptons play a suppressive role in the rates of LFV processes. Fig.~\ref{fig7}(a)-(d) also indicate that ${\rm{Br}}(h\rightarrow e\mu)$ and ${\rm{Br}}(\mu\rightarrow e\gamma)$ increase with the increasing $\tan\beta$, and decrease with the increasing $\lambda$. They affect the numerical results mainly through the mass matrices of Higgs bosons, sleptons, neutralino, so both $\tan\beta$ and $\lambda$ influence LFV in a certain degree.
 
 When $\kappa$ and $\lambda_2$ take different values, the ${\rm{Br}}(h\rightarrow e\mu)$ and ${\rm{Br}}(\mu\rightarrow e\gamma)$ with the parameters $\tan\beta^{'}$ and $g_B$ are plotted in Fig.~\ref{fig7}(e)-(f). It can be seen that ${\rm{Br}}(h\rightarrow e\mu)$ decrease slightly with the increasing $\lambda_2$, and increase slightly with the increasing $g_B$ and $\kappa$. $g_B$ affects the numerical results through the mass matrices of sleptons, Higgs bosons and neutralino, which can make small contributions to these LFV processes. Besides, the parameters $\kappa$ and $\lambda_{2}$ have also relatively small effect on ${\rm{Br}}(\mu\rightarrow e\gamma)$ and ${\rm{Br}}(h\rightarrow e\mu)$, they primarily influence the numerical results by affecting Higgs mass matrix. As can be seen from Fig.~\ref{fig7}(e), with $\tan\beta^{'}$ increasing from 1.05 to 1.15, the ${\rm{Br}}(h\rightarrow e\mu)$ can increase by almost one order of magnitude, which indicates that $\tan\beta^{'}$ is a relatively sensitive parameter and has a certain impact on the LFV.
\subsection{125 GeV Higgs LFV decay $h\rightarrow e\tau $ and $h\rightarrow \mu\tau $}    
\begin{figure}
	\setlength{\unitlength}{1mm}
	\centering
	\captionsetup{justification=raggedright,singlelinecheck=false}
	\includegraphics[width=2.8in]{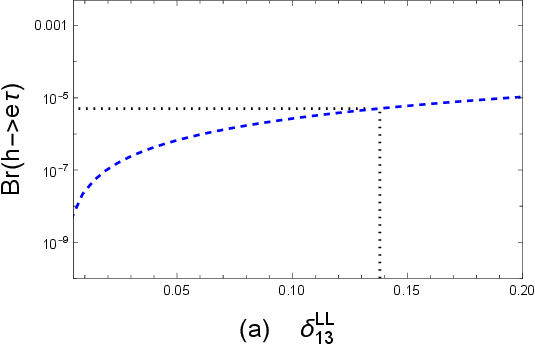}
	\vspace{0cm}
	\includegraphics[width=2.8in]{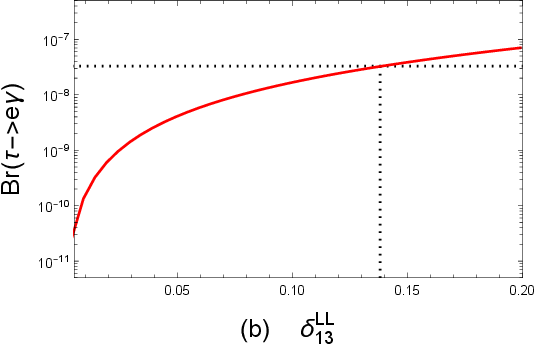}
	\vspace{0cm}
	\includegraphics[width=2.8in]{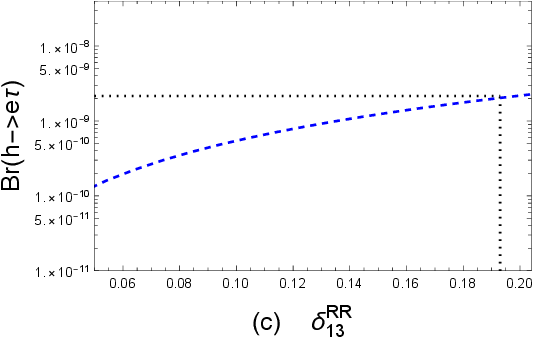}
	\vspace{0cm}
	\includegraphics[width=2.8in]{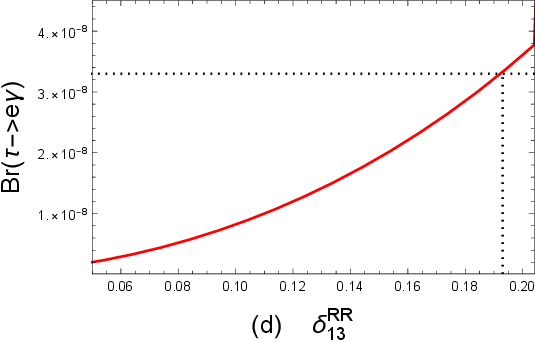}
	\vspace{0cm}
	\includegraphics[width=2.8in]{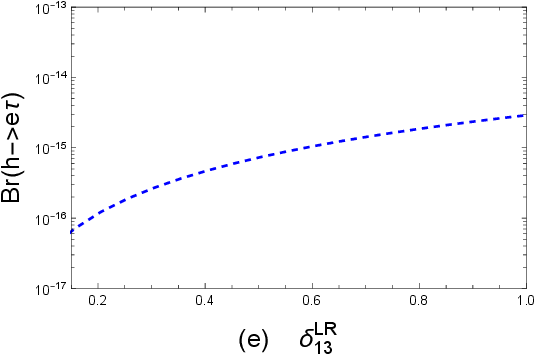}
	\vspace{0cm}
	\includegraphics[width=2.8in]{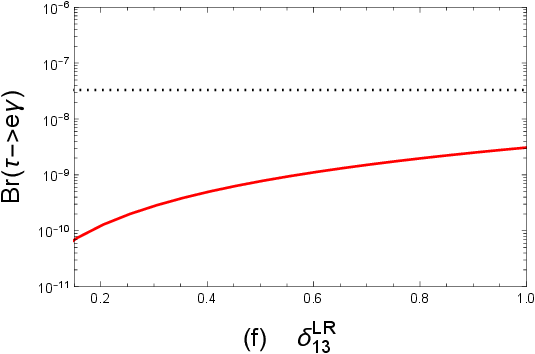}
	\vspace{0cm}
	\includegraphics[width=2.8in]{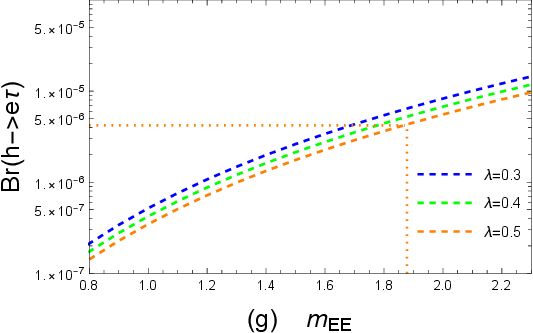}
	\vspace{0cm}
	\includegraphics[width=2.8in]{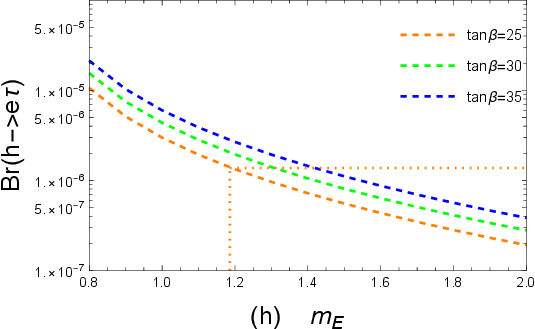}
	\vspace{0cm}
	\caption[]{{\label{fig8}} ${\rm{Br}}(h\rightarrow e\tau)$ and ${\rm{Br}}(\tau\rightarrow e\gamma)$ versus slepton flavor mixing parameters $\delta_{13}^{LL}$, $\delta_{13}^{RR}$, $\delta_{13}^{LR}$, $m_{EE}$ and $m_{E}$. Fig.~\ref{fig8}(a) and (b): $\delta_{13}^{RR}$=$\delta_{13}^{LR}$=0, Fig.~\ref{fig8}(c) and (d): $\delta_{13}^{LL}$=$\delta_{13}^{LR}$=0, Fig.~\ref{fig8}(e) and (f): $\delta_{13}^{LL}$=$\delta_{13}^{RR}$=0 .}
\end{figure}

In this section, we analyze the 125 GeV Higgs LFV decays $h\rightarrow e\tau(\mu\tau)$ in the NB-LSSM. In Fig.~\ref{fig8}, we picture ${\rm{Br}}(h\rightarrow e\tau) $ and ${\rm{Br}}(\tau \rightarrow e\gamma)$ varying with the slepton flavor mixing parameter $\delta_{13}^{XX}$ $(X=L, R) $, where the black dotted lines denote the latest experimental upper limits of  ${\rm{Br}}(\tau \rightarrow e\gamma)$. We can clearly see the influence of slepton mixing parameters $\delta^{XX}_{13}$ $(X=L, R)$ on ${\rm{Br}}(h\rightarrow e\tau) $ and ${\rm{Br}}(\tau \rightarrow e\gamma)$ from Fig.~\ref{fig8}(a) to Fig.~\ref{fig8}(f). With the increase of $\delta^{XX}_{13}$, ${\rm{Br}}(h\rightarrow e\tau) $ and ${\rm{Br}}(\tau \rightarrow e\gamma)$ all increase, so ${\rm{Br}}(h\rightarrow e\tau)$ and ${\rm{Br}}(\tau \rightarrow e\gamma) $ are proportional to the slepton mixing parameters $\delta^{XX}_{13}$. Furthermore, under the condition of other parameters being consistent, we can observe that even though the values of $\delta^{LR}_{13}$ is larger than $\delta^{LL}_{13}$ and $\delta^{RR}_{13}$, it makes the branching ratio reach very small value. Therefore, the contributions from $\delta^{LR}_{13}$ can be ignored. We also demonstrate this point in the above analytical approximate analysis in the final part of section III, and we can choose suitable parameters $\delta_{13}^{LL}$ and $\delta_{13}^{RR}$ to ensure that the experiment can detect LFV as much as possible. Considering the upper experiment limit of ${\rm{Br}}(\tau \rightarrow e\gamma)$, $\delta^{LL}_{13}$ is smaller than $0.138$ and $\delta^{RR}_{13}$ is smaller than $0.195$. Besides, ${\rm{Br}}(h\rightarrow e\tau)$ can approximate to $\mathcal{O}(10^{-6})$ in Fig.~\ref{fig8}(a), which suggests that $\delta^{LL}_{13}$ plays a crucial role in LFV.

In order to see the influence of other basic parameters on the numerical results, we set appropriate numerical values for slepton flavor mixing parameters such as $\delta_{13}^{LL}=0.06$, $\delta_{13}^{RR}=0.06$, and $\delta_{13}^{LR}=0.3$. When $\lambda=0.3(0.4,0.5)$, ${\rm{Br}}(h\rightarrow e\tau)$ versus parameter $m_{EE}$ is described by the blue, green and orange dashed lines in Fig.~\ref{fig8}(g). The figure shows that ${\rm{Br}}(h\rightarrow e\tau)$ increase quickly with the increasing $m_{EE}$ and decrease with the increasing $\lambda$. When $\tan\beta=25(30,35)$, ${\rm{Br}}(h\rightarrow e\tau)$ versus parameter $m_{E}$ is analyzed by the orange, green and blue dashed lines in Fig.~\ref{fig8}(h), which shows that ${\rm{Br}}(h\rightarrow e\tau)$ decrease quickly with the increasing $m_{E}$ and increase with the increasing $\tan\beta$. The preceding analysis demonstrates that while $m_{EE}$ and $m_{E}$ are highly sensitive parameters governing the numerical results, the contributions from $\lambda$ and $\tan\beta$ exhibit negligible effects and can therefore be ignored. Besides, the current experimental constraints of ${\rm{Br}}(h\rightarrow e\tau)$ set the limits $m_{EE}\lesssim1.9{\rm TeV}$ and $m_{E}\gtrsim1.2{\rm TeV}$. Under these experimental limits, the ${\rm{Br}}(h\rightarrow e\tau)$ can reach the order of $10^{-6}$. 

\begin{figure}
	\setlength{\unitlength}{1mm}
	\centering
	\captionsetup{justification=raggedright,singlelinecheck=false}
	\includegraphics[width=2.8in]{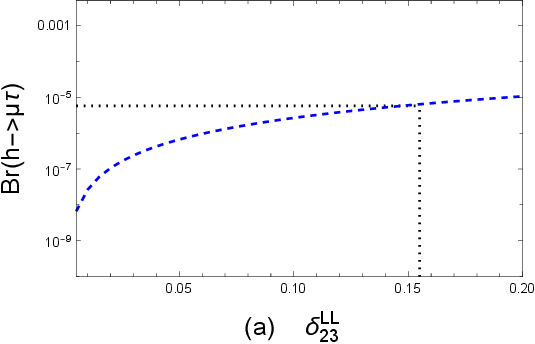}
	\vspace{0cm}
	\includegraphics[width=2.8in]{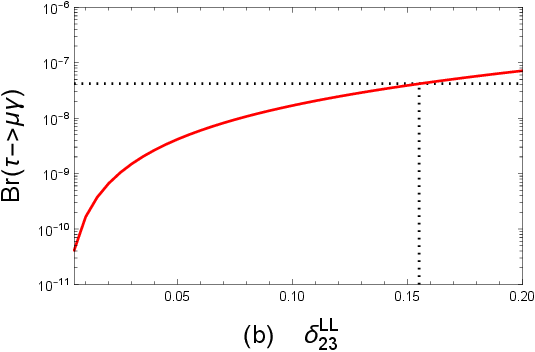}
	\vspace{0cm}
	\includegraphics[width=2.8in]{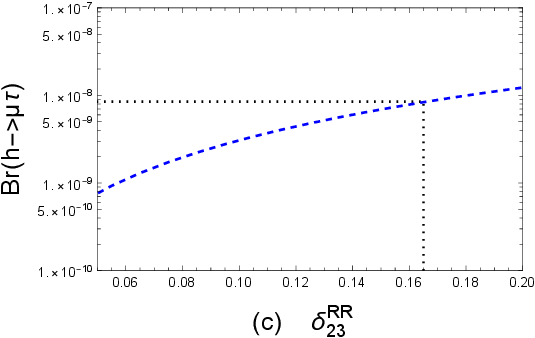}
	\vspace{0cm}
	\includegraphics[width=2.8in]{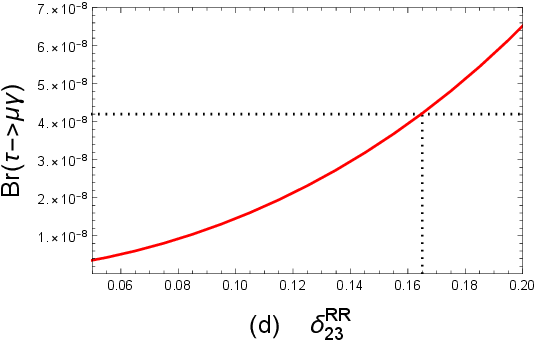}
	\vspace{0cm}
	\caption[]{{\label{fig9}} ${\rm{Br}}(h\rightarrow \mu \tau)$ and ${\rm{Br}}(\tau \rightarrow \mu\gamma)$ versus slepton flavor mixing parameters $\delta_{23}^{LL}$ and $\delta_{23}^{RR}$, Fig.~\ref{fig9}(a) and (b): $\delta_{23}^{RR}$=$\delta_{23}^{LR}$=0, Fig.~\ref{fig9}(c) and (d): $\delta_{23}^{LL}$=$\delta_{23}^{LR}$=0. }
\end{figure}

In the last, we analyze the 125 GeV Higgs LFV decays $h\rightarrow \mu\tau$ in Fig.~\ref{fig9}. We consider the restriction of the experimental upper limit of ${\rm{Br}}(\tau \rightarrow \mu\gamma)$ on process $h\rightarrow \mu\tau$ first and then discuss the influence of slepton flavor mixing parameters $\delta_{23}^{XX}$ $(X=L, R)$ on ${\rm{Br}}(h\rightarrow \mu\tau)$ and ${\rm{Br}}(\tau \rightarrow \mu\gamma)$. Fig.~\ref{fig9}(a)-(d) show us that the ${\rm{Br}}(h\rightarrow \mu\tau) $ and ${\rm{Br}}(\tau \rightarrow \mu\gamma)$ increase quickly with the increase of $\delta^{LL}_{23}$ and $\delta^{RR}_{23}$. The influence of $\delta^{LR}_{23}$ on LFV is small, so we do not talk here. Since the LFV processes are flavor dependent, the slepton flavor mixing parameters $\delta^{XX}_{23}$ $(X=L, R)$ have a large influence on ${\rm{Br}}(h\rightarrow \mu\tau)$ and ${\rm{Br}}(\tau \rightarrow \mu\gamma) $. It is obvious that ${\rm{Br}}(\tau \rightarrow \mu\gamma) $ can exceed the experimental upper limit when $\delta^{LL}_{23}$ is smaller than $0.155$ or $\delta^{RR}_{23}$ is smaller than $0.137$, and ${\rm{Br}}(h\rightarrow e\tau)$ can approximate to $\mathcal{O}(10^{-6})$ as $\delta^{LL}_{23}=0.155$ and $\delta^{RR}_{23}=\delta^{LR}_{23}=0$. Although ${\rm{Br}}(h\rightarrow \mu\tau) $ fails to reach the experimental upper limit, it is close to the experimental upper limit of ${\rm{Br}}(h\rightarrow \mu\tau)$. Therefore, $\delta^{LL}_{23}$ have a very large influence on the LFV.

\section{discussion and conclusion}

In this work, we study the LFV rare decays process $l_j\rightarrow l_i \gamma$ and 125 GeV Higgs LFV decay $h\rightarrow l_i l_j$ in the NB-LSSM. We calculate these processes in the mass eigenstate basis and the electroweak interaction basis, respectively, and the later adopt the MIA method. The numerical results from these two bases are almost identical. MIA method provides very simple and intuitive analytical formulas, and is also clear about the changes of the main parameters affecting LFV, which provides a new idea for other work of LFV in the future. 

In this study, we have performed a systematic scan of the model parameter space, considering the following experimental constraints: the Higgs boson mass within its $3\sigma$ experimental limits, Higgs signal strengths at $2\sigma$ confidence level, and the current and future experimental limits on LFV processes. This analysis has identified the viable parameter regions we choose satisfying all these constraints. Additionally, the numerical results show that the branching ratios of $h\rightarrow l_i l_j$ and $l_j\rightarrow l_i\gamma$ depend on the slepton flavor mixing parameters $\delta_{ij}^{XX}(X=L, R.$ $i, j=1, 2, 3, i \neq j)$, $m_{E}$ and $m_{EE}$ strongly. Because the LFV processes are flavor dependent, $m_{EE}$ and $\delta_{ij}^{XX}$ are present in the off-diagonal of the slepton (sneutrino) mass matrix, whose increase enlarges the branching ratios of $h\rightarrow l_i l_j$ and $l_j\rightarrow l_i\gamma$. As the diagonal terms of the slepton (sneutrino) mass matrix, the increase of $m_E$ decreases the branching ratios of $h\rightarrow l_i l_j$ and $l_j\rightarrow l_i\gamma$. Therefore, the diagonal elements of the slepton (sneutrino) mass matrix suppress the LFV effect, while the off-diagonal elements enhance the LFV effect. Under the experimental constraints of the $l_j\rightarrow l_i\gamma$, $\delta^{LL}_{12}\lesssim0.00012$, $\delta^{RR}_{12}\lesssim0.0004$, $\delta^{LR}_{12}\lesssim0.0004$, $\delta^{LL}_{13} \lesssim0.138$, $\delta^{RR}_{13} \lesssim0.195$, $\delta^{LL}_{23}\lesssim0.155$, $\delta^{RR}_{23}\lesssim0.137$, the branching ratio of $h\rightarrow e\mu$ can come up to $\mathcal{O}(10^{-14})$, and the branching ratio of $h\rightarrow e\tau(\mu\tau)$ can come up to $\mathcal{O}(10^{-6})$. The branching ratios of $h\rightarrow e\tau(\mu\tau)$ are close to the experimental upper limits of ${\rm{Br}}(h\rightarrow e\tau(\mu\tau))$, which may be detected in the near future. We also select $\tan\beta$, $\tan\beta'$, $\lambda$, $\lambda_{2}$, $\kappa$ and $g_{B}$ as variables to study Higgs LFV decays and conclude that these parameters play a role in Higgs boson decays through the trend of the lines. They affect the numerical results through the mass matrices of sleptons, Higgs bosons, sneutrino and neutralino. $\tan\beta'$ is a much sensitive parameter, whose increase leads the curve of LFV enlarge quickly. Besides, $\lambda$, $\lambda_{2}$ and $\kappa$ as new parameters of this model, can influence LFV in a certain degree. The NB-LSSM induce new sources for the LFV, which is a prospective window to search for new physics.

{\bf Acknowledgments}
This work is supported by the Major Project of National Natural Science Foundation of China (NNSFC) (No. 12235008), the National Natural Science Foundation of China (NNSFC) (No. 12075074, No. 12075073), the Natural Science Foundation of Hebei province(No.A2022201022, No. A2023201041), the Natural Science Foundation of Hebei Education Department(No. QN2022173), the Project of the China Scholarship Council (CSC) (No. 202408130113). This work is also supported by Fundação para a Ciência e a Tecnologia (FCT, Portugal) through the projects CFTP-FCT Unit UIDB/00777/2020 and UIDP/00777/2020, which are partially funded through POCTI (FEDER), COMPETE, QREN and EU.

\appendix
\section{Feynman amplitude calculation process}\label{calculation process}
In our calculation, since the mass of the external particles is much smaller than that of the internal particles, we employ the effective Lagrangian method to perform an approximate expansion of the loop integral. This approximation introduces a new physics energy scale $\Lambda$, whose magnitude is comparable to the mass scale of the internal particles. By integrating out the heavy degrees of freedom, this approach allows us to construct a low-energy effective theory containing higher-dimensional operators. The introduction of $\Lambda$ not only marks the energy scale at which new physics emerges but also ensures the convergence of the effective theory expansion. Taking Fig.~\ref{fig1}(a) as an example, we illustrate this process in detail. 
	\begin{eqnarray}
		&&i{\cal M}_{(1a)}=\langle e|\bar{u}_i(p+q)\mu^{4-D}\int\frac{d^Dk}{(2\pi)^D}
		i(H_{L}^{SF\bar{l_i} }P_L+H_{R}^{SF\bar{l_i} }P_R) 
		\frac{i}{\not{k}-m_F} i( H_{L}^{S^*l_j\bar{F}}P_L\nonumber\\&&\hspace{0.8cm}+H_{R}^{S^*l_j\bar{F}}P_R)\frac{i}{(p-k)^2-m_S^2} 
		ie(2p+q-2k)^\mu\frac{i}{((p+q-k)^2-m_S^2)} \varepsilon _{u }^{*} (q) u_{j}(p)|\mu \rangle\nonumber\\&&\hspace{0.8cm}=-\langle e|\bar{u}_i(p+q)
		\mu^{4-D}\int\frac{d^Dk}{(2\pi)^D}\frac{1}{[k^2-m_{F}^2][(k-p)^2-m_{S}^2][(k-(p+q))^2-m_{S}^2]}\nonumber\\&&\hspace{0.8cm}\times e\left(\not{k}(H_{L}^{SF\bar{l_i} } H_{R}^{S^*l_j\bar{F} } P_R+H_{R}^{SF\bar{l_i} }  H_{L}^{S^*l_j\bar{F} } P_L)+m_F(H_{L}^{SF\bar{l_i} }  H_{L}^{S^*l_j\bar{F} } P_L+H_{R}^{SF\bar{l_i} }  H_{R}^{S^*l_j\bar{F} }  P_R) \right) \nonumber\\&&\hspace{0.8cm} \times (2p+q-k)^\mu \varepsilon _{u }^{*} (q) u_{j}(p)|\mu \rangle\nonumber\\&&\hspace{0.8cm}\simeq  -\langle e|\bar{u}_i(p+q)e\mu^{4-D}\int\frac{d^Dk}{(2\pi)^D}[q^2\gamma_{\mu}\frac{k^4}{(k^2-m_F^2)(k^2-m_S^2)^4}\times\frac{1}{6} (H_{L}^{SF\bar{l_i} } H_{R}^{S^*l_j\bar{F} } P_R\nonumber\\&&\hspace{0.8cm}+H_{R}^{SF\bar{l_i} }  H_{L}^{S^*l_j\bar{F} } P_L)+(\frac{k^2}{(k^2-m_F^2)(k^2-m_S^2)^3}-\frac{1}{(k^2-m_F^2)(k^2-m_S^2)^2}) \nonumber\\&&\hspace{0.8cm}\times m_F(H_{L}^{SF\bar{l_i} }  H_{L}^{S^*l_j\bar{F} } P_L+H_{R}^{SF\bar{l_i} }  H_{R}^{S^*l_j\bar{F} }  P_R)\times i\sigma_{\mu\nu}q^{\nu}]\varepsilon _{u }^{*} (q) u_{j}(p)|\mu \rangle.
	\end{eqnarray}
	The virtual neutral fermion contributions corresponding to Fig.~\ref{fig1}(a) are deduced in the following form,
	\begin{eqnarray}
		&&iD_1^L(n)=-\frac{1}{6}\sum_{F=\chi^0_k}\sum_{S=\tilde{l}}\mu^{4-D}\int\frac{d^Dk}{(2\pi)^D}\frac{k^4}{(k^2-m_F^2)(k^2-m_S^2)^4} H_{R}^{SF\bar{l_i} }  H_{L}^{S^*l_j\bar{F} },\nonumber\\
		&&iD_2^L(n)=\sum_{F=\chi^0_k}\sum_{S=\tilde{l}}\frac{m_F}{m_{l_j}}\mu^{4-D}\int\frac{d^Dk}{(2\pi)^D}(\frac{1}{(k^2-m_F^2)(k^2-m_S^2)^2}-\frac{k^2}{(k^2-m_F^2)(k^2-m_S^2)^3}) \nonumber\\&&\hspace{0.8cm}\times H_{L}^{SF\bar{l_i} }  H_{L}^{S^*l_j\bar{F} },\nonumber\\
		&&D_{\alpha}^R(n)=D_{\alpha}^L(n)|_{L{\leftrightarrow}R},\alpha=1,2.
	\end{eqnarray}
	As a concrete example, we consider one such term: $AA$, whose explicit form is given by:
	\begin{eqnarray}
		&&AA=\mu^{4-D}\int\frac{d^Dk}{(2\pi)^D}\frac{1}{(k^2-m_F^2)(k^2-m_S^2)^2}\nonumber\\&&\hspace{0.8cm}=
		\mu^{4-D}\frac{1}{m_S^2-m_F^2}\int\frac{d^Dk}{(2\pi)^D}(\frac{1}{(k^2-m_S^2)^2}-\frac{1}{(k^2-m_S^2)(k^2-m_F^2)}.
	\end{eqnarray}
	\begin{eqnarray}
		&&\mu^{4-D}\frac{1}{m_S^2-m_F^2}\int\frac{d^Dk}{(2\pi)^D}\frac{1}{(k^2-m_S^2)^2}=\mu^{2\varepsilon  }\frac{1}{m_S^2-m_F^2}\frac{i}{(4\pi)^{2-\varepsilon}} \frac{\Gamma (\varepsilon ) }{(m_S^2)^\varepsilon  }\nonumber\\&&\hspace{0.8cm}\simeq  \frac{i}{(4\pi)^2}\frac{1}{m_S^2-m_F^2} (1+\varepsilon \ln\frac{4\pi \mu ^2}{m_S^2})(\frac{1}{\varepsilon }-\gamma_{E})\nonumber\\&&\hspace{0.8cm}\simeq\frac{1}{x_S-x_F} \frac{1}{\Lambda ^2} \frac{i}{(4\pi)^2} (\frac{1}{\varepsilon }-\gamma _E+\ln{4\pi}+\ln{x_\mu }-\ln{x_S}),\nonumber\\&&\mu^{4-D }\frac{1}{m_S^2-m_F^2}\int\frac{d^Dk}{(2\pi)^D}\frac{1}{(k^2-m_S^2)(k^2-m_F^2)}\nonumber\\&&\hspace{0.8cm}=\mu^{2\varepsilon }\frac{1}{(m_S^2-m_F^2)^2}\int\frac{d^Dk}{(2\pi)^D}(\frac{1}{(k^2-m_S^2)}-\frac{1}{(k^2-m_F^2)})\nonumber\\&&\hspace{0.8cm}\simeq \frac{1}{(m_S^2-m_F^2)^2}\frac{i}{(4\pi)^2}(\frac{1}{\varepsilon } -\gamma_{E}+1)(m_S^2(1+\varepsilon \ln\frac{4\pi \mu ^2}{m_S^2})-m_F^2(1+\varepsilon \ln\frac{4\pi \mu ^2}{m_F^2}))\nonumber\\&&\hspace{0.8cm}\simeq\frac{1}{x_S-x_F} \frac{1}{\Lambda ^2} \frac{i}{(4\pi)^2} (\frac{1}{\varepsilon }-\gamma _E+1+\ln{4\pi}+\ln{x_\mu }
		-\frac{x_S\ln x_S-x_F\ln x_F}{x_S-x_F}).
	\end{eqnarray}
	We use dimensional regularization to handle divergent terms, where $D = 4 - 2\varepsilon$, and take the limit as $D \to 4$. To obtain finite results, the divergent terms are canceled out 
	by the modified minimal subtraction($\overline{MS}$) scheme. Additionally, we introduce the new physics energy scale $\Lambda$, where $x_{\mu}=\mu^2/\Lambda^2$, $x_i=m_i^2/\Lambda^2$, with $m_i$ corresponding to the particle mass. Therefore, the $AA$ can be simplified as:
	\begin{eqnarray}
		&&AA\simeq \frac{1}{\Lambda ^2} \frac{i}{(4\pi)^2} (\frac{x_S\ln x_S-x_F\ln x_F}{(x_S-x_F)^2} -\frac{\ln x_S+1}{x_S-x_F}  )=-\frac{i}{\Lambda^2}I_3(x_F,x_S).
	\end{eqnarray}
	Then, we calculate all remaining terms following this method, thereby obtaining Eq.~(\ref{26}).

\section{One-loop functions} \label{OLF}
In this section, we give out the corresponding one-loop integral functions, which read as:
\begin{eqnarray}
&&I_1(x,y)=\frac{1}{16{\pi}^2}[-(\bigtriangleup+1+\ln{x_{\mu}})+\frac{y\ln{y}-x\ln{x}}{(y-x)}],  
\nonumber\\&&I_2(x,y)=\frac{1}{32{\pi}^2}[\frac{3+2\ln{y}}{(y-x)}-\frac{2y+4y\ln{y}}{(y-x)^2}  
+\frac{2y^2\ln{y}-2x^2\ln{x}}{(y-x)^3}],  
\nonumber\\&&I_3(x,y)=\frac{1}{16{\pi}^2}[\frac{1+\ln{y}}{(y-x)}+\frac{y\ln{y}-x\ln{x}}{(y-x)^2}],\nonumber
\end{eqnarray}
\begin{eqnarray}
&&I_4(x,y)=\frac{1}{96{\pi}^2}[\frac{11+6\ln{y}}{(y-x)}-\frac{15y+18y\ln{y}}{(y-x)^2}
+\frac{6y^2+18y^2\ln{y}}{(y-x)^3}\nonumber\\&&\hspace{2.2cm}+\frac{6x^3\ln{x}-6y^3\ln{y}}{(y-x)^4}],
\nonumber\\&&G_1(x,y,z)=\frac{1}{16{\pi}^2 }[\frac{x\ln{x}}{(x-y)(x-z)}+\frac{y\ln{y}}{(y-x)(y-z)}
+\frac{z\ln{z}}{(z-x)(z-y)}],
\nonumber\\&&G_2(x,y,z)=\frac{1}{16\pi^2} [ \frac{x^2 \ln{x}}{(y-x)(z-x)} + \frac{y^2 \ln{y}}{(x-y)(z-y)} + \frac{z^2 \ln{z}}{(x-z)(y-z)}],
\nonumber\\&&G_3(x,y,z)=\frac{1}{16\pi^2} [\frac{y^2 \ln{y} - x^2 \ln{x}}{(y-x)^2} + \frac{y + 2y \ln{y}}{(x-y)}],
\nonumber\\&&G_5(x,y,z,t)=\frac{1}{16{\pi}^2 {\Lambda}^4}[\frac{x\ln{x}}{(y-x)(x-z)(x-t)}+\frac{y\ln{y}}{(x-y)(y-z)(y-t)}
\nonumber\\&&\hspace{2.2cm}+\frac{z\ln{z}}{(x-z)(z-y)(z-t)} +\frac{t\ln{t}}{(x-t)(z-y)(t-y)(t-z)}],
\nonumber\\&&G_6(x,y,z,t)=\frac{1}{16\pi^2{\Lambda}^2} [\frac{x^2 \ln{x}}{(y-x)(z-x)(t-x)} +\frac{y^2 \ln{y}}{(x-y)(z-y)(t-y)}\nonumber\\&&\hspace{2.2cm}+\frac{z^2 \ln{z}}{(x-z)(y-z)(t-z)}+\frac{t^2 \ln{t}}{(x-t)(t-y)(t-z)}],
\nonumber\\&&G_7(x,y,z,t)=\frac{1}{16\pi^2{\Lambda}^4} [\frac{-t^2 \ln{t}}{(t-x)^2(t-y)(t-z)}+\frac{y^2 \ln{y}}{(x-y)^2(y-z)(t-y)}\nonumber\\&&\hspace{2.2cm}-\frac{z^2 \ln{z}}{(x-z)^2(y-z)(t-z)} +\frac{x(x^3+2tyz-xty-tz-yz) \ln{x}}{(t-x)^2(x-y)^2(x-z)^2}\nonumber\\&&\hspace{2.2cm}-\frac{1}{(t-x)(t-y)(x-y)(x-z)(y-z)}],\nonumber
\end{eqnarray}
\begin{eqnarray}
&&G_8(x,y,z,t)=\frac{1}{16\pi^2{\Lambda}^6} [\frac{-t \ln{t}}{(t-x)^2(t-y)(t-z)}+\frac{y \ln{y}}{(x-y)^2(y-z)(t-y)}\nonumber\\&&\hspace{2.2cm}+\frac{z \ln{z}}{(x-z)^2(y-z)(z-t)} +\frac{(-2{x^3}+{x^2}y+{x^2}z+{x^2}t-tyz) \ln{x}}{(x-y)^2}\nonumber\\&&\hspace{2.2cm}+\frac{1}{(t-x)(x-y)(x-z)}],\nonumber\\&&G_{9}(x, y, z, t, n)=[G_{5}(x, y, z, t)-G_{5}(x, y, z, n)]\frac{1}{(t-n)\Lambda^2}, \nonumber\\
&&G_{10}(x, y, z, t, n)=[G_{6}(x, y, z, t)-G_{6}(x, y, z, n)]\frac{1}{(t-n)\Lambda^2}, \
\end{eqnarray}
with $\Delta=\frac{1}{\epsilon}-\gamma_{E}+\ln{4\pi}$.

\section{The relevant Feynman rules in the electroweak interaction basis} \label{Feynman rules}
\begin{figure}
	\includegraphics[width=12.5cm]{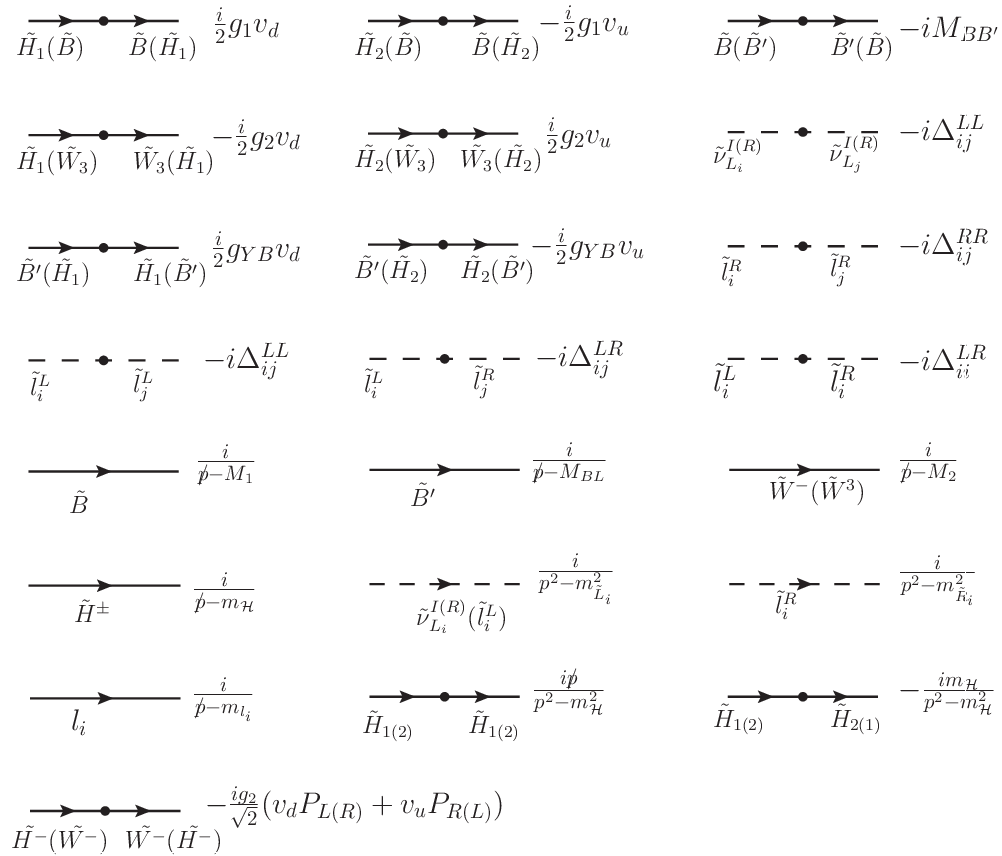}
	\caption{Feyman rules for the relevant insertions and  relevant propagators}
	\label{1234}
\end{figure}

The relevant Feynman rules for the present computation are collected in Fig.~\ref{1234}.

\section{The contributions from Feynman diagrams with the MIA method}\label{MR}
To save space in the text, we use some symbols to represent the coupling vertices associated with the Higgs boson, as follows:
\begin{eqnarray}
	&&B=(-g_{YB}g_{B}+g_{1}^{2}+g_{YB}^{2}+g_{2}^{2})v_dZ_{11}^{H}+g_{YB}g_{B}+g_{1}^{2}+g_{YB}^{2}
	+g_{2}^{2})v_uZ_{12}^{H}\nonumber\\
	&&~~~~~~-2(g_{YB}g_{B}+g_{B}^{2})(-v_{\bar{\eta } } Z_{14}^{H}+v_{\eta} Z_{13}^{H}),\\
	&&Q=\frac{1}{2} (-\sqrt{2} T_{e,ii}Z _{11}^{H} +\lambda \upsilon_s Y_{e,ii} Z _{12}^{H}),\\ &&D=\frac{1}{2} (-\sqrt{2} T_{e,ij}Z _{11}^{H}),\
\end{eqnarray}
\begin{eqnarray}
&&N=-\frac{1}{\sqrt{2}} Y_{e,jj}Z _{11}^{H}.\
\end{eqnarray}

The one-loop contributions from Fig.~\ref{fig3}(1a,1b)
\begin{eqnarray}
	&&K'_{R}(1a,1b)_{\text{MIA}}=\frac{1}{2}g_{2}Y_{e,jj}\Delta_{ij}^{LL}[G_{5}(x_{\tilde{\nu}^{I}_{L_{j}}},x_{\tilde{\nu}^{I}_{L_{i}}},x_{2},x_{\mathcal{H} })m_{\mathcal{H} } (-\frac{1}{\sqrt{2}}g_2 Z_{12}^{H}) M_2\nonumber\\
	&&~~~~~~~~~~~~+G_{6}(x_{\tilde{\nu}^{I}_{L_{j}}},x_{\tilde{\nu}^{I}_{L_{i}}},x_{2},x_{\mathcal{H} }) (-\frac{1}{\sqrt{2}}g_2 Z_{11}^{H})]\nonumber\\
	&&~~~~~~~~~~~~+\frac{1}{2}g_{2}Y_{e,jj}\Delta_{ij}^{LL}[G_{5}(x_{\tilde{\nu}^{R}_{L_{j}}},x_{\tilde{\nu}^{R}_{L_{i}}},x_{2},x_{\mathcal{H} })m_{\mathcal{H} } (-\frac{1}{\sqrt{2}}g_2 Z_{12}^{H}) M_2\nonumber\\
	&&~~~~~~~~~~~~+G_{6}(x_{\tilde{\nu}^{R}_{L_{j}}},x_{\tilde{\nu}^{R}_{L_{i}}},x_{2},x_{\mathcal{H} }) (-\frac{1}{\sqrt{2}}g_2 Z_{11}^{H})].
\end{eqnarray}
The one-loop contributions from Fig.~\ref{fig3}(2a,2b)
\begin{eqnarray}
	&&K'_{R}(2a,2b)_{\text{MIA}}=-\frac{1}{8}g_2Y_{e,jj}B\Delta_{ij}^{LL}
	[G_{7}(x_{\tilde{\nu}^{I}_{L_{j}}},x_{\tilde{\nu}^{I}_{L_{i}}},x_{2},x_{\mathcal{H} })
	(-g_2 \frac{v_d}{\sqrt{2}})\nonumber\\
	&&~~~~~~~~~~~~~~~~~
	+G_{8}(x_{\tilde{\nu}^{I}_{L_{j}}},x_{\tilde{\nu}^{I}_{L_{i}}},x_{2},x_{\mathcal{H} })
	(-g_2 \frac{v_u}{\sqrt{2}})m_{\mathcal{H} } M_2 ]\nonumber\\
	&&~~~~~~~~~~~~~~~~~-\frac{1}{8}g_2Y_{e,jj}B\Delta_{ij}^{LL}
	[G_{7}(x_{\tilde{\nu}^{R}_{L_{j}}},x_{\tilde{\nu}^{R}_{L_{i}}},x_{2},x_{\mathcal{H} })
	(-g_2 \frac{v_d}{\sqrt{2}})\nonumber\\
	&&~~~~~~~~~~~~~~~~~
	+G_{8}(x_{\tilde{\nu}^{R}_{L_{j}}},x_{\tilde{\nu}^{R}_{L_{i}}},x_{2},x_{\mathcal{H} })
	(-g_2 \frac{v_u}{\sqrt{2}})m_{\mathcal{H} } M_2 ].
\end{eqnarray}
The one-loop contributions from Fig.~\ref{fig3}(2c,2d)
\begin{eqnarray}
	&&K'_{R}(2c,2d)_{\text{MIA}}=-\frac{1}{8} g_2Y_{e,jj}B\Delta_{ij}^{LL}
	[G_{7}(x_{\tilde{\nu}^{I}_{L_{i}}},x_{\tilde{\nu}^{I}_{L_{j}}},x_{2},x_{\mathcal{H} })
	(-g_2 \frac{v_d}{\sqrt{2}})\nonumber\\
	&&~~~~~~~~~~~~~~~~~+G_{8}(x_{\tilde{\nu}^{I}_{L_{i}}},x_{\tilde{\nu}^{I}_{L_{j}}},x_{2},x_{\mathcal{H} })
	(-g_2 \frac{v_u}{\sqrt{2}})m_{\mathcal{H} } M_2 ]\nonumber\\
	&&~~~~~~~~~~~~~~~~~-\frac{1}{8} g_2Y_{e,jj}B\Delta_{ij}^{LL}
	[G_{7}(x_{\tilde{\nu}^{R}_{L_{i}}},x_{\tilde{\nu}^{R}_{L_{j}}},x_{2},x_{\mathcal{H} })
	(-g_2 \frac{v_d}{\sqrt{2}})\nonumber\\
	&&~~~~~~~~~~~~~~~~~+G_{8}(x_{\tilde{\nu}^{R}_{L_{i}}},x_{\tilde{\nu}^{R}_{L_{j}}},x_{2},x_{\mathcal{H} })
	(-g_2 \frac{v_u}{\sqrt{2}})m_{\mathcal{H} } M_2 ].	
\end{eqnarray}
The one-loop contributions from Fig.~\ref{fig3}(3a-3f)
\begin{eqnarray}
	&&K'_{R}(3a-3f)_{\text{MIA}}=\frac{\sqrt{2} }{4}g_{1}^{2} Y_{e,jj}Z_{12}^{H} \Delta_{ij}^{LL}
	G_{5}(x_{\tilde{l}^{L}_{j}},x_{\tilde{l}^{L}_{i}},x_{1},x_{\mathcal{H} })m_{\mathcal{H} } M_1\nonumber\\
	&&~~~~~~~~~~~~~~~~~+\frac{\sqrt{2}}{4}g_{YB}(g_{YB}+g_{B}) Y_{e,jj}Z_{12}^{H} \Delta_{ij}^{LL}
	G_{5}(x_{\tilde{l}^{L}_{j}},x_{\tilde{l}^{L}_{i}},x_{BL},x_{\mathcal{H} })m_{\mathcal{H} } M_{BL}\nonumber\\
	&&~~~~~~~~~~~~~~~~~-\frac{\sqrt{2}}{4}g_{2}^{2} Y_{e,jj}Z_{12}^{H} \Delta_{ij}^{LL}
	G_{5}(x_{\tilde{l}^{L}_{j}},x_{\tilde{l}^{L}_{i}},x_{2},x_{\mathcal{H} })m_{\mathcal{H} } M_2\nonumber
\end{eqnarray}
\begin{eqnarray}
	&&~~~~~~~~~~~~~~~~~+\frac{\sqrt{2}}{4}g_{1}^{2} Y_{e,jj}Z_{12}^{H} \Delta_{ij}^{LL}
	[G_{6}(x_{\tilde{l}^{L}_{j}},x_{\tilde{l}^{L}_{i}},x_{1},x_{\mathcal{H} })-\frac{m_{l_j}^2}{2}G_{7}(x_{\mathcal{H} },x_{1},x_{\tilde{l}^{L}_{j}},x_{\tilde{l}^{L}_{i}})]\nonumber\\
	&&~~~~~~~~~~~~~~~~~-\frac{\sqrt{2}}{4}g_{2}^{2} Y_{e,jj}Z_{12}^{H} \Delta_{ij}^{LL}
	[G_{6}(x_{\tilde{l}^{L}_{j}},x_{\tilde{l}^{L}_{i}},x_{2},x_{\mathcal{H} })-\frac{m_{l_j}^2}{2}G_{7}(x_{\mathcal{H} },x_{2},x_{\tilde{l}^{L}_{j}},x_{\tilde{l}^{L}_{i}})]\nonumber\\
	&&~~~~~~~~~~~~~~~~~+\frac{\sqrt{2}}{4}g_{YB}(g_{YB}+g_{B}) Y_{e,jj}Z_{12}^{H} \Delta_{ij}^{LL}
	[G_{6}(x_{\tilde{l}^{L}_{j}},x_{\tilde{l}^{L}_{i}},x_{BL},x_{\mathcal{H} })\nonumber\\
	&&~~~~~~~~~~~~~~~~~-\frac{m_{l_j}^2}{2}G_{7}(x_{\mathcal{H} },x_{BL},x_{\tilde{l}^{L}_{j}},x_{\tilde{l}^{L}_{i}})].
\end{eqnarray}
The one-loop contributions from Fig.~\ref{fig3}(3g-3j)
\begin{eqnarray}
	&&K'_{L}(3g-3j)_{\text{MIA}}=-\frac{\sqrt{2} }{2}g_{1}^{2} Y_{e,jj}Z_{12}^{H} \Delta_{ij}^{RR}
	G_{5}(x_{\tilde{l}^{R}_{j}},x_{\tilde{l}^{R}_{i}},x_{1},x_{\mathcal{H} })m_{\mathcal{H} } M_1\nonumber\\
	&&~~~~~~~~~~~~~~~~~-\frac{\sqrt{2}}{4} g_{YB}(2g_{YB}+g_{B}) Y_{e,jj}Z_{12}^{H} \Delta_{ij}^{RR}
	G_{5}(x_{\tilde{l}^{R}_{j}},x_{\tilde{l}^{R}_{i}}x_{BL},x_{\mathcal{H} })m_{\mathcal{H} } M_{BL}\nonumber\\
	&&~~~~~~~~~~~~~~~~~-\frac{\sqrt{2}}{2} g_{1}^{2} Y_{e,jj}Z_{12}^{H} \Delta_{ij}^{RR}
	[G_{6}(x_{\tilde{l}^{R}_{j}},x_{\tilde{l}^{R}_{i}},x_{1},x_{\mathcal{H} })-\frac{m_{l_j}^2}{2}G_{7}(x_{\mathcal{H} },x_{1},x_{\tilde{R}_{j}},x_{\tilde{R}_{i}})]\nonumber\\
	&&~~~~~~~~~~~~~~~~~-\frac{\sqrt{2}}{4} g_{YB}(2g_{YB}+g_{B}) Y_{e,jj}Z_{12}^{H} \Delta_{ij}^{RR}
	[G_{6}(x_{\tilde{l}^{R}_{j}},x_{\tilde{l}^{R}_{i}},x_{BL},x_{\mathcal{H} })\nonumber\\
	&&~~~~~~~~~~~~~~~~~-\frac{m_{l_j}^2}{2}G_{7}(x_{\mathcal{H} },x_{BL},x_{\tilde{l}^{R}_{j}},x_{\tilde{l}^{R}_{i}})].\
\end{eqnarray}
The one-loop contributions from Fig.~\ref{fig3}(3k-3p)
\begin{eqnarray}
	&&K'_{L}(3k-3p)_{\text{MIA}}=\frac{\sqrt{2} }{4}g_{1}^{2} Y_{e,ii}Z_{12}^{H} \Delta_{ij}^{LL}
	G_{5}(x_{\tilde{l}^{L}_{j}},x_{\tilde{l}^{L}_{i}},x_{1},x_{\mathcal{H} })m_{\mathcal{H} } M_1\nonumber\\
	&&~~~~~~~~~~~~~~~~~+\frac{\sqrt{2}}{4}g_{YB}(g_{YB}+g_{B}) Y_{e,ii}Z_{12}^{H} \Delta_{ij}^{LL}
	G_{5}(x_{\tilde{l}^{L}_{j}},x_{\tilde{l}^{L}_{i}}x_{BL},x_{\mathcal{H} })m_{\mathcal{H} } M_{BL}\nonumber\\
	&&~~~~~~~~~~~~~~~~~-\frac{\sqrt{2}}{4}g_{2}^{2} Y_{e,ii}Z_{12}^{H} \Delta_{ij}^{LL}
	G_{5}(x_{\tilde{l}^{L}_{j}},x_{\tilde{l}^{L}_{i}},x_{2},x_{{\mathcal{H} }})m_{\mathcal{H} } M_2\nonumber\\
	&&~~~~~~~~~~~~~~~~~+\frac{\sqrt{2}}{4}g_{1}^{2} Y_{e,ii}Z_{12}^{H} \Delta_{ij}^{LL}
	[G_{6}(x_{\tilde{l}^{L}_{j}},x_{\tilde{l}^{L}_{i}},x_{1},x_{\mathcal{H} })-\frac{m_{l_j}^2}{2}G_{7}(x_{\mathcal{H} },x_{1},x_{\tilde{l}^{L}_{j}},x_{\tilde{l}^{L}_{i}})]\nonumber\\
	&&~~~~~~~~~~~~~~~~~-\frac{\sqrt{2}}{4}g_{2}^{2} Y_{e,ii}Z_{12}^{H} \Delta_{ij}^{LL}
	[G_{6}(x_{\tilde{l}^{L}_{j}},x_{\tilde{l}^{L}_{i}},x_{2},x_{\mathcal{H} })-\frac{m_{l_j}^2}{2}G_{7}(x_{\mathcal{H} },x_{2},x_{\tilde{l}^{L}_{j}},x_{\tilde{l}^{L}_{i}})]\nonumber\\
	&&~~~~~~~~~~~~~~~~~+\frac{\sqrt{2}}{4}g_{YB}(g_{YB}+g_{B}) Y_{e,ii}Z_{12}^{H} \Delta_{ij}^{LL}
	[G_{6}(x_{\tilde{l}^{L}_{j}},x_{\tilde{l}^{L}_{i}},x_{BL},x_{\mathcal{H} })\nonumber\\
	&&~~~~~~~~~~~~~~~~~-\frac{m_{l_j}^2}{2}G_{7}(x_{\mathcal{H} },x_{BL},x_{\tilde{l}^{L}_{j}},x_{\tilde{l}^{L}_{i}})].
\end{eqnarray}
The one-loop contributions from Fig.~\ref{fig3}(3q-3t)
\begin{eqnarray}
    &&K'_{R}(3q-3t)_{\text{MIA}}=-\frac{\sqrt{2} }{2}g_{1}^{2} Y_{e,ii}Z_{12}^{H} \Delta_{ij}^{RR}
    G_{5}(x_{\tilde{l}^{R}_{j}},x_{\tilde{l}^{R}_{i}},x_{1},x_{\mathcal{H} })m_{\mathcal{H} } M_1\nonumber\\
    &&~~~~~~~~~~~~~~~~~~~-\frac{\sqrt{2}}{4}g_{YB}(2g_{YB}+g_{B}) Y_{e,ii}Z_{12}^{H} \Delta_{ij}^{RR}G_{5}(x_{\tilde{l}^{R}_{j}},x_{\tilde{l}^{R}_{i}},x_{BL},x_{\mathcal{H} })m_{\mathcal{H} } M_{BL}\nonumber\\
    &&~~~~~~~~~~~~~~~~~~~-\frac{\sqrt{2}}{2}g_{1}^{2} Y_{e,ii}Z_{12}^{H} \Delta_{ij}^{RR}
    [G_{6}(x_{\tilde{l}^{R}_{j}},x_{\tilde{l}^{R}_{i}},x_{1},x_{\mathcal{H} })-\frac{m_{l_j}^2}{2}G_{7}(x_{\mathcal{H} },x_{1},x_{\tilde{l}^{R}_{j}},x_{\tilde{l}^{R}_{i}})]\nonumber\\
    &&~~~~~~~~~~~~~~~~~~~-\frac{\sqrt{2}}{4}g_{YB}(2g_{YB}+g_{B}) Y_{e,ii}Z_{12}^{H} \Delta_{ij}^{RR}[G_{6}(x_{\tilde{l}^{R}_{j}},x_{\tilde{l}^{R}_{j}},x_{\tilde{l}^{R}_{i}},x_{BL},x_{\mathcal{H} })\nonumber\\
    &&~~~~~~~~~~~~~~~~~~~-\frac{m_{l_j}^2}{2}G_{7}(x_{\mathcal{H} },x_{BL},x_{\tilde{l}^{R}_{j}},x_{\tilde{l}^{R}_{i}})].\
\end{eqnarray}
The one-loop contributions from Fig.~\ref{fig3}(4a-4d)
\begin{eqnarray}
	&&K'_{L}(4a,4b)_{\text{MIA}}=\frac{1}{\Lambda^2}g_{1}^{2}D G_{1}(x_{\tilde{l}^{L}_{j}},x_{\tilde{l}^{R}_{i}},x_{1})M_1\nonumber\\
	&&~~~~~~~~~~~~~~+\frac{1}{2\Lambda^2}(2g_{YB}+g_{B})(g_{YB}+g_{B})D G_{1}(x_{\tilde{l}^{L}_{j}},x_{\tilde{l}^{R}_{i}},x_{BL})M_{BL},\nonumber\\
	&&K'_{R}(4c,4d)_{\text{MIA}}=\frac{1}{\Lambda^2}g_{1}^{2}D G_{1}(x_{\tilde{l}^{R}_{j}},x_{\tilde{l}^{L}_{i}},x_{1})M_1\nonumber\\
	&&~~~~~~~~~~~~~~+\frac{1}{2\Lambda^2}(2g_{YB}+g_{B})(g_{YB}+g_{B})D G_{1}(x_{\tilde{l}^{R}_{j}},x_{\tilde{l}^{L}_{i}},x_{BL})M_{BL}.
\end{eqnarray}
The one-loop contributions from Fig.~\ref{fig3}(5a-5d)
\begin{eqnarray}
    &&K'_{L}(5a,5b)_{\text{MIA}}=g_{1}(g_{YB}+g_{B})D M_{BB'}[G_{6}(x_{\tilde{l}^{R}_{i}},x_{\tilde{l}^{L}_{j}},x_{BL},x_1)\nonumber\\
    &&~~~~~~~~~~~~+G_{5}(x_{\tilde{l}^{R}_{i}},x_{\tilde{l}^{L}_{j}},x_{BL},x_1) M_{BL} M_1]\nonumber\\
    &&~~~~~~~~~~~~+\frac{1}{2}(2g_{YB}+g_{B})g_1 D M_{BB'}[G_{6}(x_{\tilde{l}^{R}_{i}},x_{\tilde{l}^{L}_{j}},x_{BL},x_1)\nonumber\\
    &&~~~~~~~~~~~~+G_{5}(x_{\tilde{l}^{R}_{i}},x_{\tilde{l}^{L}_{j}},x_{BL},x_1) M_{BL} M_1],\nonumber\\
    &&K'_{R}(5c,5d)_{\text{MIA}}=g_{1}(g_{YB}+g_{B})D M_{BB'}[G_{6}(x_{\tilde{l}^{R}_{j}},x_{\tilde{l}^{L}_{i}},x_{BL},x_1)\nonumber\\
    &&~~~~~~~~~~~~+G_{5}x_{\tilde{l}^{R}_{j}},x_{\tilde{l}^{L}_{i}},x_{BL},x_1) M_{BL} M_1]\nonumber\\
    &&~~~~~~~~~~~~+\frac{1}{2}(2g_{YB}+g_{B})g_1 D M_{BB'}[G_{6}(x_{\tilde{l}^{R}_{j}},x_{\tilde{l}^{L}_{i}},x_{BL},x_1)\nonumber\\
    &&~~~~~~~~~~~~+G_{5}(x_{\tilde{l}^{R}_{j}},x_{\tilde{l}^{L}_{i}},x_{BL},x_1) M_{BL} M_1].
\end{eqnarray}
The one-loop contributions from Fig.~\ref{fig3}(6a-6h)
\begin{eqnarray}
	&&K'_{L}(6a,6b)_{\text{MIA}}=g_{1}^{2} P \Delta_{ij}^{RR}G_{5}(x_{\tilde{l}^{R}_{j}},x_{\tilde{l}^{R}_{i}},x_{\tilde{l}^{L}_{j}},x_1) M_1 \nonumber\\
	&&~~~~~~~~~~~~+\frac{1}{2}(2g_{YB}+g_B)(g_{YB}+g_B) P \Delta_{ij}^{RR}G_{5}(x_{\tilde{l}^{R}_{j}},x_{\tilde{l}^{R}_{i}},x_{\tilde{l}^{L}_{j}},x_{BL}) M_{BL},\nonumber\\
	&&K'_{L}(6c,6d)_{\text{MIA}}=g_{1}^{2} Q \Delta_{ij}^{LL}G_{5}(x_{\tilde{l}^{L}_{i}},x_{\tilde{l}^{L}_{j}},x_{\tilde{l}^{R}_{i}},x_1) M_1 \nonumber\\
	&&~~~~~~~~~~~~+\frac{1}{2}(2g_{YB}+g_B)(g_{YB}+g_B) Q \Delta_{ij}^{LL}G_{5}(x_{\tilde{l}^{L}_{i}},x_{\tilde{l}^{L}_{j}},x_{\tilde{l}^{R}_{i}},x_{BL}) M_{BL},\nonumber\\
	&&K'_{R}(6e,6f)_{\text{MIA}}=g_{1}^{2} Q \Delta_{ij}^{RR}G_{5}(x_{\tilde{l}^{R}_{j}},x_{\tilde{l}^{R}_{i}},x_{\tilde{l}^{L}_{i}},x_1) M_1 \nonumber\\
    &&~~~~~~~~~~~~+\frac{1}{2}(2g_{YB}+g_B)(g_{YB}+g_B) Q \Delta_{ij}^{RR}G_{5}(x_{\tilde{l}^{R}_{j}},x_{\tilde{l}^{R}_{i}},x_{\tilde{l}^{L}_{i}},x_{BL}) M_{BL},\nonumber\\
    &&K'_{R}(6g,6h)_{\text{MIA}}=g_{1}^{2} P \Delta_{ij}^{LL}G_{5}(x_{\tilde{l}^{L}_{i}},x_{\tilde{l}^{L}_{j}},x_{\tilde{l}^{R}_{j}},x_1) M_1 \nonumber\\
    &&~~~~~~~~~~~~+\frac{1}{2}(2g_{YB}+g_B)(g_{YB}+g_B) P \Delta_{ij}^{LL}G_{5}(x_{\tilde{l}^{L}_{i}},x_{\tilde{l}^{L}_{j}},x_{\tilde{l}^{R}_{j}},x_{BL}) M_{BL}.	
\end{eqnarray}

The one-loop contributions from Fig.~\ref{fig3}(7a-7h)
\begin{eqnarray}
&&K'_{R}(7a,7b)_{\text{MIA}}=g_{1}(g_{YB}+g_{B})Q \Delta_{ij}^{RR} M_{BB'}[G_{10}(x_{\tilde{l}^{R}_{i}},x_{\tilde{l}^{R}_{j}},x_{\tilde{l}^{L}_{i}},x_{BL},x_1)\nonumber\\
&&~~~~~~~~~~+G_{9}(x_{\tilde{l}^{R}_{i}},x_{\tilde{l}^{R}_{j}},x_{\tilde{l}^{L}_{i}},x_{BL},x_1) M_{BL} M_1]\nonumber\\
&&~~~~~~~~~~+\frac{1}{2}(2g_{YB}+g_{B})g_1 Q \Delta_{ij}^{RR} M_{BB'}[G_{10}(x_{\tilde{l}^{R}_{i}},x_{\tilde{l}^{R}_{j}},x_{\tilde{l}^{L}_{i}},x_{BL},x_1)\nonumber\\
&&~~~~~~~~~~+G_{9}(x_{\tilde{l}^{R}_{i}},x_{\tilde{l}^{R}_{j}},x_{\tilde{l}^{L}_{i}},x_{BL},x_1) M_{BL} M_1],\nonumber
\end{eqnarray}
\begin{eqnarray}
&&K'_{L}(7c,7d)_{\text{MIA}}=g_{1}(g_{YB}+g_{B})P \Delta_{ij}^{RR} M_{BB'}[G_{10}(x_{\tilde{l}^{R}_{i}},x_{\tilde{l}^{R}_{j}},x_{\tilde{l}^{L}_{i}},x_{BL},x_1)\nonumber\\
&&~~~~~~~~~~+G_{9}(x_{\tilde{l}^{R}_{i}},x_{\tilde{l}^{R}_{j}},x_{\tilde{l}^{L}_{i}},x_{BL},x_1) M_{BL} M_1]\nonumber\\
&&~~~~~~~~~~+\frac{1}{2}(2g_{YB}+g_{B})g_1 P \Delta_{ij}^{RR} M_{BB'}[G_{10}(x_{\tilde{l}^{R}_{i}},x_{\tilde{l}^{R}_{j}},x_{\tilde{l}^{L}_{i}},x_{BL},x_1)\nonumber\\
&&~~~~~~~~~~+G_{9}(x_{\tilde{l}^{R}_{i}},x_{\tilde{l}^{R}_{j}},x_{\tilde{l}^{L}_{i}},x_{BL},x_1) M_{BL} M_1],\nonumber\\ 
&&K'_{L}(7e,7f)_{\text{MIA}}=g_{1}(g_{YB}+g_{B})Q \Delta_{ij}^{LL} M_{BB'}[G_{10}(x_{\tilde{l}^{L}_{i}},x_{\tilde{l}^{L}_{j}},x_{\tilde{l}^{R}_{i}},x_{BL},x_1)\nonumber\\&&~~~~~~~~~~+G_{9}(x_{\tilde{l}^{L}_{i}},x_{\tilde{l}^{L}_{j}},x_{\tilde{l}^{R}_{i}},x_{BL},x_1) M_{BL} M_1]\nonumber\\
&&~~~~~~~~~~+\frac{1}{2}(2g_{YB}+g_{B})g_1 Q \Delta_{ij}^{LL} M_{BB'}[G_{10}(x_{\tilde{l}^{L}_{i}},x_{\tilde{l}^{L}_{j}},x_{\tilde{l}^{R}_{i}},x_{BL},x_1)\nonumber\\
&&~~~~~~~~~~+G_{9}(x_{\tilde{l}^{L}_{i}},x_{\tilde{l}^{L}_{j}},x_{\tilde{l}^{R}_{i}},x_{BL},x_1) M_{BL} M_1],\nonumber
\end{eqnarray}
\begin{eqnarray}
&&K'_{R}(7g,7h)_{\text{MIA}}=g_{1}(g_{YB}+g_{B})P \Delta_{ij}^{LL} M_{BB'}[G_{10}(x_{\tilde{l}^{L}_{i}},x_{\tilde{l}^{L}_{j}},x_{\tilde{l}^{R}_{i}},x_{BL},x_1)\nonumber\\
&&~~~~~~~~~~+G_{9}(x_{\tilde{l}^{L}_{i}},x_{\tilde{l}^{L}_{j}},x_{\tilde{l}^{R}_{i}},x_{BL},x_1) M_{BL} M_1]\nonumber\\
&&~~~~~~~~~~+\frac{1}{2}(2g_{YB}+g_{B})g_1 P \Delta_{ij}^{LL} M_{BB'}[G_{10}(x_{\tilde{l}^{L}_{i}},x_{\tilde{l}^{L}_{j}},x_{\tilde{l}^{R}_{i}},x_{BL},x_1)\nonumber\\
&&~~~~~~~~~~+G_{9}(x_{\tilde{l}^{L}_{i}},x_{\tilde{l}^{L}_{j}},x_{\tilde{l}^{R}_{i}},x_{BL},x_1) M_{BL} M_1]. 
\end{eqnarray}
The one-loop contributions from Fig.~\ref{fig4}(1a-1d)
\begin{eqnarray}
	&&K'_{R}(1a-1d)_{\text{MIA}}=\frac{1}{2\sqrt{2} }Z_{11}^{H}g_{2}Y_{e,jj}^2\Delta_{ij}^{LL}[(-g_2 \frac{v_d}{\sqrt{2}}) (G_{6}(x_{\tilde{\nu}^{I}_{L_{j}}},x_{\tilde{\nu}^{I}_{L_{i}}},x_{2},x_{\mathcal{H} })\nonumber\\
	&&~~~~~~~~~~~~~~~~+G_{6}(x_{\tilde{\nu}^{R}_{L_{j}}},x_{\tilde{\nu}^{R}_{L_{i}}},x_{2},x_{\mathcal{H} }))+m_{\mathcal{H} } M_2(-g_2 \frac{v_u}{\sqrt{2}})(G_{5}(x_{\tilde{\nu}^{I}_{L_{j}}},x_{\tilde{\nu}^{I}_{L_{i}}},x_{2},x_{\mathcal{H} })\nonumber\\
	&&~~~~~~~~~~~~~~~~+G_{5}(x_{\tilde{\nu}^{R}_{L_{j}}},x_{\tilde{\nu}^{R}_{L_{i}}},x_{2},x_{\mathcal{H} }))]\frac{m_{l_j} }{m_{l_j}^{2}-m_{l_i}^{2}}. \
\end{eqnarray}
The one-loop contributions from Fig.~\ref{fig4}(2a-2h)
\begin{eqnarray}
	&&K'_{L}(2a-2d)_{\text{MIA}}=\frac{m_{l_j} }{(m_{l_j}^{2}-m_{l_i}^{2}) \Lambda^2}[-g_{1}^{2}\Delta_{ij}^{LR} N G_{1}(x_{\tilde{l}^{L}_{i}},x_{\tilde{l}^{R}_{j}},x_{1}) M_1 \nonumber\\
	&&~~~~~~~~~~~~-\frac{1}{2}(2g_{YB}+g_{B})(g_{YB}+g_{B})\Delta_{ij}^{LR} N G_{1}(x_{\tilde{l}^{L}_{i}},x_{\tilde{l}^{R}_{j}},x_{BL}) M_{BL}], \nonumber\\
	&&K'_{R}(2e-2h)_{\text{MIA}}=\frac{m_{l_j} }{(m_{l_j}^{2}-m_{l_i}^{2}) \Lambda^2}[-g_{1}^{2}\Delta_{ij}^{LR} N G_{1}(x_{\tilde{l}^{R}_{j}},x_{\tilde{l}^{L}_{i}},x_{1}) M_1 \nonumber\\
	&&~~~~~~~~~~~~-\frac{1}{2}(2g_{YB}+g_{B})(g_{YB}+g_{B})\Delta_{ij}^{LR} N G_{1}(x_{\tilde{l}^{R}_{j}},x_{\tilde{l}^{L}_{i}},x_{BL}) M_{BL}] .
\end{eqnarray}
The one-loop contributions from Fig.~\ref{fig4}(3a-3h)
\begin{eqnarray}
	&&K'_{L}(3a-3d)_{\text{MIA}}=\frac{m_{l_j} }{m_{l_j}^{2}-m_{l_i}^{2}}[-g_1 (g_{YB}+g_{B}) M_{BB^{'}} N \Delta_{ij}^{LR}\big(G_{6}(x_{\tilde{l}^{L}_{j}},x_{\tilde{l}^{R}_{i}},x_{1},x_{BL})\nonumber\\
	&&~~~~~~~~~~~~+G_{5}(x_{\tilde{l}^{L}_{j}},x_{\tilde{l}^{R}_{i}},x_{1},x_{BL})M_1 M_{BL}\big)-\frac{1}{2}g_1(2g_{YB}+g_{B})M_{BB^{'}} N \Delta_{ij}^{LR}(G_{6}(x_{\tilde{l}^{L}_{j}},x_{\tilde{l}^{R}_{i}},x_{1},x_{BL})\nonumber\\
	&&~~~~~~~~~~~~+G_{5}(x_{\tilde{l}^{L}_{j}},x_{\tilde{l}^{R}_{i}},x_{1},x_{BL})M_1 M_{BL})],\nonumber\\
	&&K'_{R}(3e-3h)_{\text{MIA}}=\frac{m_{l_j} }{m_{l_j}^{2}-m_{l_i}^{2}}[-\frac{1}{2}g_1(2g_{YB}+g_{B})M_{BB^{'}} N \Delta_{ij}^{LR}(G_{6}(x_{\tilde{l}^{L}_{i}},x_{\tilde{l}^{R}_{j}},x_{1},x_{BL})\nonumber\\
	&&~~~~~~~~~~~~+G_{5}(x_{\tilde{l}^{L}_{i}},x_{\tilde{l}^{R}_{j}},x_{1},x_{BL})M_1 M_{BL})-g_1 (g_{YB}+g_{B}) M_{BB^{'}} N \Delta_{ij}^{LR}(G_{6}(x_{\tilde{l}^{L}_{i}},x_{\tilde{l}^{R}_{j}},x_{1},x_{BL})\nonumber\\
	&&~~~~~~~~~~~~+G_{5}(x_{\tilde{l}^{L}_{i}},x_{\tilde{l}^{R}_{j}},x_{1},x_{BL})M_1 M_{BL})].
\end{eqnarray}

The one-loop contributions from Fig.~\ref{fig4}(4a-4l)
\begin{eqnarray}
	&&K'_{R}(4a-4l)_{\text{MIA}}=\frac{m_{l_j} }{m_{l_j}^{2}-m_{l_i}^{2}}[g_{1}^{2} N Y_{e,jj} \frac{\sqrt{2}}{4}\Delta_{ij}^{LL} v_{d} G_{6}(x_{\tilde{l}^{L}_{i}},x_{\tilde{l}^{L}_{j}},x_{1},x_{\mathcal{H} })\nonumber\\
	&&~~~~~~~~~~~~-g_{1}^{2} N Y_{e,jj} \frac{\sqrt{2}}{4}\Delta_{ij}^{LL} v_{u} G_{5}(x_{\tilde{l}^{L}_{i}},x_{\tilde{l}^{L}_{j}},x_{1},x_{\mathcal{H} })m_{\mathcal{H} } M_{1} \nonumber\\
	&&~~~~~~~~~~~~+\frac{\sqrt{2}}{4} g_{YB}(g_{YB}+g_{B})N \Delta_{ij}^{LL} Y_{e,jj}v_{d} G_{6}(x_{\tilde{l}^{L}_{j}},x_{\tilde{l}^{L}_{i}},x_{BL},x_{\mathcal{H} }) \nonumber\\
	&&~~~~~~~~~~~~-\frac{\sqrt{2}}{4} g_{YB}(g_{YB}+g_{B})N \Delta_{ij}^{LL} Y_{e,jj}v_{u} G_{5}(x_{\tilde{l}^{L}_{i}},x_{\tilde{l}^{L}_{j}},x_{BL},x_{\mathcal{H} })m_{\mathcal{H} } M_{BL} \nonumber\\
	&&~~~~~~~~~~~~-g_{2}^{2} \frac{\sqrt{2}}{4}  N \Delta_{ij}^{LL} Y_{e,jj}v_{d} G_{6}(x_{\tilde{l}^{L}_{j}},x_{\tilde{l}^{L}_{i}},x_{2},x_{\mathcal{H} }) \nonumber\\
	&&~~~~~~~~~~~~-g_{2}^{2} \frac{\sqrt{2}}{4}  N \Delta_{ij}^{LL} Y_{e,jj}v_{u} G_{5}(x_{\tilde{l}^{L}_{j}},x_{\tilde{l}^{L}_{i}},x_{\mathcal{H} },x_{2})m_{\mathcal{H} } M_2 ].
\end{eqnarray}
The one-loop contributions from Fig.~\ref{fig4}(4m-4t)
\begin{eqnarray}
	&&K'_{L}(4m-4t)_{\text{MIA}}=\frac{m_{l_j} }{m_{l_j}^{2}-m_{l_i}^{2}}[g_{1}^{2} N Y_{e,jj} \frac{\sqrt{2}}{2}\Delta_{ij}^{RR} v_{d} G_{6}(x_{\tilde{l}^{R}_{i}},x_{\tilde{l}^{R}_{j}},x_{1},x_{\mathcal{H} })\nonumber\\
	&&~~~~~~~~~~~~~~~~~~-g_{1}^{2} N Y_{e,jj} \frac{\sqrt{2}}{2}\Delta_{ij}^{RR} v_{u} G_{5}(x_{\tilde{l}^{R}_{i}},x_{\tilde{l}^{R}_{j}},x_{1},x_{\mathcal{H} })m_{\mathcal{H} } M_{1} \nonumber\\
	&&~~~~~~~~~~~~~~~~~~-g_{YB}(2g_{YB}+g_{B})\frac{\sqrt{2}}{4} N \Delta_{ij}^{RR} Y_{e,jj}v_{u} G_{5}(x_{\tilde{l}^{R}_{i}},x_{\tilde{l}^{R}_{j}},x_{BL},x_{\mathcal{H} })m_{\mathcal{H} } M_{BL} \nonumber\\
	&&~~~~~~~~~~~~~~~~~~+g_{YB}(2g_{YB}+g_{B})\frac{\sqrt{2}}{4} N \Delta_{ij}^{RR} Y_{e,jj}v_{d} G_{6}(x_{\tilde{l}^{R}_{j}},x_{\tilde{l}^{R}_{i}},x_{BL},x_{\mathcal{H} })] .\
\end{eqnarray}

 \end{document}